%% file: pTZ_for_PDFs.tex
\documentclass[a4paper,10pt]{article}
\pdfoutput=1
\usepackage{jheppub}

\usepackage[T1]{fontenc} % if needed
\usepackage{bbold}
\usepackage{amsmath}
\usepackage{empheq}
\usepackage{booktabs}
\usepackage{slashed}
\usepackage{url}
\usepackage{tabularx}

\newcommand{\rd}{{\rm d}}

\newcommand{\re}{{\rm e}}

\newcommand{\as}{\alpha_s}

\usepackage{xspace}

\newcommand{\rad}{\textsc{RadISH}\xspace}
\newcommand{\nnlojet}{\textsc{NNLOjet}\xspace}

\newcommand{\eq}[1]{eq.~(\ref{#1})}
\newcommand{\csq}{\chi^2}

\newcommand{\pt}{p_T}
\newcommand{\yll}{\ensuremath{y_{\ell\ell}}\xspace}
\newcommand{\mll}{\ensuremath{m_{\ell\ell}}\xspace}
\newcommand{\ptll}{\ensuremath{\pt^{\ell\ell}}\xspace}

\newcommand{\ptmll}{\ensuremath{p_{T,{\min}}^{\ell\ell}}\xspace}

\def\({\left(} 
\def\){\right)}

\newcommand{\beq}{\begin{eqnarray}}
\newcommand{\eeq}{\end{eqnarray}}

\usepackage[dvipsnames]{xcolor}
\usepackage[normalem]{ulem}

\counterwithin{figure}{section}
\counterwithin{table}{section}
\counterwithin{equation}{section}

\title{\boldmath Impact of $Z$-boson transverse-momentum resummation on
  PDF determination}

\author[a]{Juan M.~Cruz-Martinez,}
\author[b]{Emanuele R.~Nocera,}
\author[c]{Luca Rottoli,}
\author[b]{and Paolo Torrielli.}

\affiliation[a]{Departamento de Fisica Atomica, Molecular y Nuclear,
  Facultad de Fisica,\\ Universidad de Sevilla E-41080 Sevilla, Spain}
\affiliation[b]{Dipartimento di Fisica, Universit\`a di Torino, and INFN,
  Sezione di Torino,\\Via P. Giuria 1, I-10125 Torino, Italy}
\affiliation[c]{Dipartimento di Fisica, Universit\`a di Milano-Bicocca, and
  INFN, Sezione di Milano-Bicocca,\\Piazza della Scienza 3, I-20126 Milano,
  Italy}

\emailAdd{jcruz@us.es}
\emailAdd{emanueleroberto.nocera@unito.it}
\emailAdd{luca.rottoli@unimb.it}
\emailAdd{paolo.torrielli@unito.it}

\abstract{We study the impact of small-transverse-momentum resummation for
  neutral-current Drell--Yan lepton-pair production on the determination of
  collinear parton distribution functions (PDFs). We focus on measurements of
  the $Z$-boson transverse-momentum spectrum performed by ATLAS and CMS at the
  LHC at centre-of-mass energies of 8 and 13~TeV, and include them in PDF fits
  based on the NNPDF methodology. Theoretical predictions are computed at
  next-to-next-to-leading order (NNLO) in perturbative QCD and are supplemented
  with small-transverse-momentum resummation corrections at
  next-to-next-to-next-to-leading logarithmic accuracy obtained with \rad.
  Missing higher-order uncertainties are accounted for through a theory
  covariance matrix constructed from renormalisation- and factorisation-scale
  variations. We first revisit the treatment of the 8~TeV data, replacing the
  fixed-order predictions used in previous analyses with numerically stable
  NNLO calculations, and removing the additional numerical uncertainties
  introduced in earlier fits. We then assess the impact of the 13~TeV
  measurements, of resummation corrections, and of progressively lowering the
  minimum cut on the dilepton transverse momentum. We find that resummation
  improves the description of the $Z$ boson transverse-momentum data and is
  essential for ensuring the overall consistency of the PDF fits. Nevertheless,
  it does not conclusively support extending the fitted kinematic region to
  transverse momenta below a few tens of GeV. The impact of resummation on PDFs
  is moderate, leading primarily to a stabilisation of the gluon PDF in a
  kinematic region of relevance for LHC phenomenology. Finally, we comment on
  the fact that the treatment of correlations among theoretical uncertainties
  may play a central role in PDF fits to measurements with percent- and
  sub-percent-level precision.}

\begin{document} 
\maketitle
\flushbottom

\input{sec-introduction.tex}

\input{sec-input.tex}
\input{sec-results.tex}
\input{sec-conclusions.tex}
\input{acknowledgements.tex}

\appendix

\input{app-weighted_fit.tex}

\bibliographystyle{JHEP}
\bibliography{pTZ_for_PDFs}

\end{document}

%% file: sec-introduction.tex
\section{Introduction}
\label{sec:introduction}

The results obtained during Run II have established the Large Hadron Collider
(LHC) as a precision machine that has exceeded all expectations, rivalling
lepton colliders in the measurement of electroweak (EW) precision observables.
In this context, the production of lepton pairs through the Drell--Yan (DY)
mechanism plays a central role. It allows for a precise measurement of Standard
Model (SM) parameters such as the masses and widths of the $Z$ and $W$
bosons~\cite{LHCb:2021bjt,ATLAS:2024erm,CMS:2024lrd,LHCb:2025nob,LHCb:2025msn},
the EW mixing
angle~\cite{ATLAS:2015ihy,LHCb:2015jyu,CMS:2018ktx,LHCb:2024ygc,CMS:2024ony},
and the strong coupling $\alpha_s$~\cite{ATLAS:2023lhg}.
Furthermore, the astonishing precision of the available data makes the DY
process central in the calibration of Monte Carlo simulation codes and provides
important constraints in the determination of parton distribution functions
(PDFs). For these reasons, as well as for the relative simplicity of the DY
process compared to other LHC processes, theoretical predictions for
lepton-pair production have reached a very high level of precision.
Differential predictions for Born-level observables are known at
next-to-next-to-next-to-leading order (N$^3$LO)
QCD~\cite{Chen:2021vtu,Chen:2022cgv,Chen:2022lwc,Chen:2022lpw,Neumann:2022lft,
  Campbell:2023lcy}. Next-to-leading order (NLO) EW~\cite{Dittmaier:2001ay,
  Baur:2004ig,Zykunov:2006yb,Arbuzov:2005dd,CarloniCalame:2006zq,Baur:2001ze,
  Zykunov:2005tc,CarloniCalame:2007cd,Arbuzov:2007db,Dittmaier:2009cr} and
mixed next-to-next-to-leading order (NNLO) QCD$\times$EW
effects~\cite{Buonocore:2021rxx,Bonciani:2021zzf,Buccioni:2022kgy,
  Armadillo:2024ncf} are also available for the full off-shell process.

Kinematic distributions involving the lepton pair are particularly clean
observables, as they are sensitive to accompanying hadronic activity only
through kinematic recoil. Among them, the transverse momentum $\ptll$ of the
dilepton system in neutral-current (NC) DY production plays a pivotal role.
Experimentally, it is measured with uncertainties well below the $1\%$ level 
when normalised spectra are considered, see {\it e.g.}~\cite{ATLAS:2019zci}. 
On the theory side, predictions for this distribution are available at NNLO in
QCD~\cite{Boughezal:2015dva,Gehrmann-DeRidder:2015wbt,Boughezal:2015ded,
Boughezal:2016dtm,Boughezal:2016isb,Gehrmann-DeRidder:2016cdi,
Gehrmann-DeRidder:2016jns,Gauld:2017tww,Gehrmann-DeRidder:2017mvr,
Gauld:2021pkr}, and have been matched to calculations that include
the all-order resummation of logarithmically enhanced contributions at low
$\ptll$ up to N$^3$LL$^\prime$ or partial N$^4$LL
accuracy~\cite{Neumann:2022lft,Camarda:2023dqn,Campbell:2023lcy,
Moos:2023yfa,Billis:2024dqq}. In some cases, these calculations also
incorporate the interplay between QCD and EW corrections~\cite{Cieri:2018sfk,
Autieri:2023xme,Buonocore:2024xmy}. At large $\ptll$, the distribution
becomes increasingly sensitive to large EW Sudakov logarithms, making the
inclusion of higher-order EW corrections essential~\cite{Boughezal:2017nla}.

The transverse-momentum distribution in NC DY provides important constraints on
the extraction of PDFs~\cite{Boughezal:2017nla}. While inclusive measurements
maximally control quark flavour separation, the $\ptll$ spectrum is sensitive to
the gluon PDF at intermediate-to-large values of Bjorken-$x$, thus providing a
complementary constraint to the gluon PDF with respect to single-inclusive jet
and top-quark pair production data~\cite{Nocera:2017zge,Cridge:2023ozx}.
Notably, the inclusion of $\ptll$ experimental data directly shapes
the PDF constraints relevant for Higgs physics, since Higgs production at the
LHC probes the same region in $x$. For these reasons, the $\ptll$ distribution
is included in all the most recent global PDF
determinations~\cite{Ablat:2025gbp,Bailey:2020ooq,NNPDF:2021njg}.

The high precision of the available $\ptll$ data, however, makes
their inclusion in PDF determinations particularly delicate. An accurate
description of this distribution requires a consistent treatment of several
effects of comparable importance, such as the perturbative accuracy of the
theoretical predictions, the size and correlation of missing higher-order QCD
corrections, the impact of all-order resummation at low $\ptll$, and
that of EW logarithms at high $\ptll$. Moreover, the
small experimental uncertainties require excellent numerical precision in the
codes used to compute theoretical predictions, as the quality of the fit may
be sensitive to residual numerical instabilities.

In this paper we quantitatively assess the impact of these effects on an
accurate and precise determination of PDFs.  We do so by considering the
inclusion of a specific set of high-precision experimental measurements of NC
DY data in a PDF determination carried out with the NNPDF
methodology~\cite{NNPDF:2021njg}. We focus on the ATLAS and CMS LHC measurements
at a centre-of-mass energy of~8~TeV~\cite{ATLAS:2015iiu,CMS:2015hyl}
and 13~TeV~\cite{ATLAS:2019zci,CMS:2022ubq}. The 8~TeV data set was already
included in NNPDF4.0~\cite{NNPDF:2021njg}, whereas the 13~TeV data set is new
and has not been included in any collinear PDF determination to date. These
measurements cover a range in $\ptll$ extending down to a few GeV.
The data at low $\ptll$ are currently discarded from PDF determinations
because fixed-order theory is unreliable in that region. For the first time,
we study the impact of small-$\ptll$ resummation on the determination
of PDFs, and analyse the possibility of including data at very low values of
$\ptll$. In the case of the 8~TeV data, we concurrently revisit the
precision of the perturbative computation and the treatment of uncertainties.
In all cases, we make sure that our analysis is not spoiled by EW Sudakov logarithms,
thanks to a judicious choice of large-$\ptll$ cuts.

The paper is organised as follows. In Sect.~\ref{sec:input}, we review the
experimental, theoretical, and methodological input to our study, focusing in
particular on general differences with respect to the reference NNPDF4.0
analysis~\cite{Ball:2025xgq}, and on the treatment of $Z$-boson transverse
momentum measurements. In Sect.~\ref{sec:results}, we outline the details
of the PDF fits that we perform, and we discuss the results incrementally:
first, the impact of theoretical and computational improvements; second,
the impact of the new 13~TeV $Z$-boson transverse momentum measurements; and
third the effect of resummation corrections, both as a function of the data
set and as a function of the minimum cut on $\ptll$.
In Sect.~\ref{sec:conclusions}, we finally summarise our conclusions. The
paper is complemented by Appendix~\ref{app:weighted_fit}, in which we
further discuss the consistency and the compatibility of the ATLAS 13~TeV
measurement~\cite{ATLAS:2019zci} with the rest of the NNPDF4.0 data set.

%% file: sec-input.tex
\section{Experimental, theoretical, and methodological input}
\label{sec:input}

In this section, we review the experimental, theoretical, and methodological
ingredients of our study. We first describe the experimental measurements
considered, with particular emphasis on the $Z$-boson transverse-momentum
distribution. We then outline the theoretical framework used to compute the
corresponding predictions, and finally summarise the NNPDF fitting methodology
within which our study is carried out.

\subsection{Experimental data}
\label{subsec:data}

The baseline data set considered in this work is the one used to determine
the NNPDF4.0 parton set together with the strong coupling
$\alpha_s$~\cite{Ball:2025xgq}. It incorporates a wide range of measurements
for lepton-nucleon, neutrino-nucleus,
proton-nucleus, and proton-(anti)proton scattering processes, specifically:
NC and charged-current (CC) deep-inelastic scattering (DIS), from both 
fixed-target~\cite{NewMuon:1996uwk,NewMuon:1996fwh,Whitlow:1991uw,
  BCDMS:1989qop,CHORUS:2005cpn,NuTeV:2001dfo,Mason:2006qa}
and collider~\cite{H1:2015ubc,H1:2018flt} experiments; NC and CC DY,
from both fixed-target~\cite{Moreno:1990sf,NuSea:2003qoe,NuSea:2001idv,
  SeaQuest:2021zxb} and collider~\cite{CDF:2010vek,D0:2007djv,D0:2014kma,
  ATLAS:2014ape,ATLAS:2013xny,ATLAS:2011qdp,ATLAS:2016nqi,CMS:2012ivw,
  CMS:2013pzl,CMS:2013zfg,LHCb:2012gii,LHCb:2015okr,ATLAS:2017rue,
  ATLAS:2016gic,CMS:2015hyl,LHCb:2015kwa,LHCb:2015mad,LHCb:2016zpq,
  ATLAS:2016fij,LHCb:2016fbk} experiments, including DY in association with
an extra jet~\cite{ATLAS:2014jkm,CMS:2013wql,CMS:2018dxg,ATLAS:2017irc,
  ATLAS:2015iiu,CMS:2015hyl}; top-quark
pair~\cite{CMS:2017zpm,ATLAS:2014nxi,Spannagel:2016cqt,
  ATLAS:2020aln,CMS:2015yky,ATLAS:2015lsn,ATLAS:2016pal,CMS:2015rld,
  CMS:2017iqf,CMS:2018htd,CMS:2018adi},
single-inclusive jet~\cite{ATLAS:2017kux,CMS:2016lna},
dijet~\cite{ATLAS:2013jmu,CMS:2012ftr}
prompt-photon~\cite{ATLAS:2017nah},
and single top-quark~\cite{ATLAS:2014sxe,CMS:2012xhh,ATLAS:2017rso,CMS:2014mgj,
  ATLAS:2016qhd,CMS:2016lel} production from the LHC.

Among the aforementioned collider NC DY measurements, those of the
$Z$-boson production cross section differential in $\ptll$
are of particular relevance to this work. Such measurements have
been performed at the LHC by ATLAS~\cite{ATLAS:2015iiu} and
CMS~\cite{CMS:2015hyl} at a centre-of-mass energy of 8~TeV.
Concerning ATLAS, the data set corresponds to an integrated
luminosity of 20.3~fb$^{-1}$. We consider two subsets of the data.
The first consists of the absolute cross section, differential in
the dilepton transverse momentum $\ptll$ and rapidity
$\yll$, and covers the invariant-mass region around the $Z$-boson
peak. The second consists of the cross section differential in
$\ptll$ and the dilepton invariant mass $\mll$, and covers
the off-peak invariant-mass regions.
We always consider measurements in which electron and muon yields are
combined to maximise their statistical precision, which globally amounts
to few percent. Concerning CMS, the measurement corresponds to an integrated
luminosity of 19.7~fb$^{-1}$.
We consider the absolute distribution differential in $\yll$
and $\ptll$, which covers the peak invariant-mass region.
In this case, the $Z$ boson is identified by its decay into a pair of
muons only. The total uncertainty amounts to a few percent as for the
ATLAS measurement.

In this paper, for the first time, we also study the analogous
ATLAS~\cite{ATLAS:2019zci} and CMS~\cite{CMS:2022ubq} measurements
performed at a centre-of-mass energy of 13~TeV.
Concerning ATLAS, the measurement corresponds to an integrated luminosity
of 36.1~fb$^{-1}$, stemming from the (relatively small) fraction of LHC
RunII events collected in 2015 and 2016.
We consider the normalised distribution differential in $\ptll$, which
covers the $Z$-peak invariant-mass region, and combines electron and muon
decay channels.
The data are unprecedentedly precise, with uncertainties at the permille level
at low values of $\ptll$.
As for CMS, the measurement corresponds to a very similar integrated luminosity
of 36.3~fb$^{-1}$. Here we consider the absolute distribution, differential
in the invariant mass and transverse momentum of the combined electron and
muon pairs, which covers both the $Z$-peak and off-peak invariant-mass regions.
The total uncertainty is of the order of 1\% or less.

In all cases, we account for the correlations among experimental
uncertainties using the information provided by the corresponding
measurements.
Kinematic cuts are as in~\cite{Ball:2025xgq}. Concerning the ATLAS and CMS
$Z$-boson transverse-momentum  measurements, we apply two sets of kinematic
cuts to the data.
The first one, at small $\ptll$, tests the region where resummation
corrections are large. We start by selecting $\ptll\geq 30$~GeV, as required
in~\cite{NNPDF:2017mvq,NNPDF:2021njg}. A central goal of this work,
however, is to assess the quality of the data–theory comparison and of the
resulting PDF fits as this cut is progressively lowered, as we shall do in
Sect.~\ref{subsec:res}. The second kinematic cut, at large $\ptll$, is designed
to exclude the region where EW corrections may become large.
As in~\cite{NNPDF:2017mvq,NNPDF:2021njg},
we require $\ptll\leq 150$~GeV for both the 8 and 13~TeV data.
The dependence of theoretical predictions and PDFs on this cut was studied
for the ATLAS and CMS 8~TeV measurements in~\cite{Boughezal:2017nla}.
A similar behaviour is observed for the 13~TeV data (see for instance Fig.~6
of~\cite{ATLAS:2019zci}).

\subsection{Theoretical predictions}
\label{subsec:theory}

We compute theoretical predictions at NNLO accuracy in the strong coupling.
In contrast to~\cite{Ball:2025xgq}, higher-order corrections are no longer
included by means of $K$-factors.
They are instead directly incorporated into interpolation grids --- specifically
in the {\sc PineAPPL}~\cite{Carrazza:2020gss,Jezo:2026adf,
christopher_schwan_2025_15635174} format --- which allow us to store
precomputed weights for the perturbative part of the cross section.
The numerical evaluation of hadronic cross sections then becomes extremely
fast, as it reduces to a weighted sum of interpolating functions evaluated
on grid nodes.
This feature is essential to make PDF fits computationally feasible, as the
fitting procedure requires convolving PDFs with partonic cross sections
a large number of times throughout parameter optimisation.
Even though we do not expect that replacing
$K$-factors with the exact computation is going to affect PDFs in a significant
way~\cite{Cruz-Martinez:2025ffa}, we note that in principle $K$-factors
suffer from the limitations of being insensitive to partonic channel
decomposition and of being mildly PDF-dependent. Using exact NNLO grids
removes these limitations. This is a first difference, and
a potential improvement,
in comparison to~\cite{Ball:2025xgq} and all
previous NNPDF determinations~\cite{Boughezal:2017nla,NNPDF:2017mvq,
NNPDF:2021njg}.

In addition to the use of exact NNLO grids, there are two further differences
in comparison to~\cite{Ball:2025xgq}. For single-inclusive jet and dijet
production, we use the interpolation grids
computed in~\cite{Britzger:2022lbf}, which include full-colour NNLO corrections
to dijet production~\cite{Chen:2022tpk}
in contrast to the leading-colour approximation used in previous NNPDF fits.
For DY measurements, we use the interpolation grids computed
in~\cite{Cruz-Martinez:2025ffa},
which take the transverse mass of the vector boson as the central
renormalisation and factorisation scale rather than the boson mass.
This choice is consistent with the scale adopted in this paper for the
calculations of the Z-boson transverse momentum distributions.
The software used to compute the grid weights depends on the process:
{\sc yadism}~\cite{Candido:2024rkr,barontini_2026_18758473} for NC
and CC DIS; {\sc MATRIX}~\cite{Grazzini:2017mhc} for top-quark pair production;
{\sc Vrap}~\cite{Anastasiou:2003ds,Barontini:2023vmr} for fixed-target DY;
and \nnlojet~\cite{NNLOJET:2025rno} for all other processes.

We supplement NNLO computations with an account of missing
higher-order uncertainties as in~\cite{Ball:2025xgq}. Concerning $Z$-boson
transverse-momentum measurements, we crucially incorporate small-$\ptll$
resummation corrections. In the following we discuss the details of fixed-order,
resummed, and missing higher-order computations, focusing on the $Z$-boson
transverse-momentum measurements.

\subsubsection{Fixed-order: \nnlojet}
\label{subsubsec:NNLOjet}

We compute predictions for the $Z$-boson transverse momentum at NNLO accuracy
in the strong coupling with \nnlojet~\cite{NNLOJET:2025rno}. We set
central factorisation and renormalisation scales to
$\mu_F=\mu_R=\sqrt{(\ptll)^2+\mll^2}$,
where $\ptll$ and $\mll$ are the transverse momentum and the invariant mass
of the final-state lepton pair. To accomplish the cancellation of infrared
singularities among real and virtual contributions at NLO and NNLO,
\nnlojet implements the antenna subtraction
method~\cite{Gehrmann-DeRidder:2005btv,Daleo:2006xa,Currie:2013vh}, which
derives the subtraction terms starting from a colour-ordered decomposition of
all real-radiation contributions.
Colour ordering makes it possible to associate each unresolved real-radiation
configuration with a pair of hard radiator partons. Each such configuration
is then described by an antenna function consisting of the two hard radiators
and the unresolved radiation emitted between them.
The antenna subtraction method is a local subtraction scheme, meaning that the
subtraction terms cancel the singular limits of the real-radiation matrix
elements point-by-point in phase space. As a result, no slicing cutoff is
required. This feature enables the efficient computation of numerically stable,
fully differential predictions in the presence of non-trivial fiducial cuts. 

In contrast, the NNLO predictions for the ATLAS and CMS measurements of the
$Z$-boson transverse-momentum distribution at
8~TeV~\cite{ATLAS:2015iiu,CMS:2015hyl}, included in the reference NNPDF
analysis~\cite{Ball:2025xgq}, were provided by the authors of
Refs.~\cite{Boughezal:2015ded,Boughezal:2016dtm,Boughezal:2016isb}
and computed using the $N$-jettiness non-local subtraction
technique~\cite{Boughezal:2015dva,Boughezal:2015eha,Gaunt:2015pea}.
In its standard formulation, this method amounts to slicing the radiative
phase space by means of a $\tau_N^{\rm cut}$ cut in the $N$-jettiness
event-shape variable $\tau_N$. For $\tau_N>\tau_N^{\rm cut}$, one uses the
exact NLO calculation for NC DY with one extra resolved jet;
in the $\tau_N<\tau_N^{\rm cut}$ region, the cross section is approximated
by the fixed-order expansion of the $\tau_N$ resummation, as obtained in
soft-collinear effective theory (SCET) \cite{Bauer:2000yr,Bauer:2001yt},
which induces power corrections in $\tau_N^{\rm cut}$.
Therefore, the numerical accuracy of this method is limited not only by
Monte Carlo statistics, but also by the level of control one has on power
corrections.

Because of this fact, in~\cite{Boughezal:2017nla} it was observed that NNLO
corrections, implemented as multiplicative
$K$-factors~\cite{Boughezal:2015ded,Boughezal:2016isb},
were numerically unstable\footnote{We note that an improved treatment of
$\tau_N^{\rm cut}$ power corrections is implemented in recent versions of the
{\sc MCFM} code~\cite{Neumann:2022lft}.}.
Fluctuations exceeded the nominal Monte Carlo uncertainty
of the computation, and turned out to be larger than the typical statistical
uncertainty of the data. It was therefore concluded that the nominal
uncertainty of the computation was underestimated. In order to get a more
realistic estimate, in Ref.~\cite{Carrazza:2017bjw} an ensemble of neural
network fits to the NNLO/NLO cross-section ratio was performed, as a function
of the transverse momentum and rapidity of the $Z$ boson, for each of the
rapidity bins of the ATLAS and CMS measurements. The one-sigma uncertainty of
the fit, induced by the point-to-point fluctuations of the NNLO predictions,
was estimated to be approximately $1\%$. To account for this effect, an
additional $1\%$ uncorrelated uncertainty was then assigned to the experimental
data. In Sect.~\ref{subsec:baseline}, we revisit the need for this additional
uncertainty in light of the improved numerical precision provided by the
\nnlojet computation.

\subsubsection{Resummation: \rad}
\label{subsubsec:resumm}

We account for resummation corrections by means of the \rad
framework~\cite{Monni:2016ktx,Bizon:2017rah,Monni:2019yyr,Re:2021con,
  Buonocore:2024xmy}, which is designed for the all-order resummation of global,
recursively infrared- and collinear- (rIRC) safe~\cite{Banfi:2001bz,
  Banfi:2003je,Banfi:2004yd} observables $V$ that vanish away from the Sudakov
limit. In our case, the key observable is the normalised transverse momentum
$V=\ptll/\mll$ of the lepton pair in NC DY production, which has
been studied extensively in the \rad framework~\cite{Bizon:2018foh,
  Bizon:2019zgf,Re:2021con,Chen:2022cgv,Chen:2022lpw,Rottoli:2023xdc,
  Torrielli:2023tiz,Buonocore:2024xmy}. The resummation formalism is applied in
momentum space, where all expressions are cast as integrals that are finite in
four-space dimensions, and evaluated by means of Monte Carlo techniques.
Schematically, the \rad prediction for the cumulative cross section for
observable $V$ being smaller than value $v$, indicated by $\Sigma(v)$, can be
written fully differentially with respect to the Born phase-space variables
$\Phi_B$ as a convolution integral over the transverse momentum $k_{t1}$ of the
hardest radiation in the ensemble of emitted soft/collinear QCD partons,
\beq
\label{eq:master-kt-space}
\frac{\rd\Sigma(v)}{\rd\Phi_B}
& = &
\int
\frac{\rd k_{t1}}{k_{t1}} \,
{\cal L}(k_{t1}) \, \re^{-R(k_{t1})} \, {\cal F}(v, \Phi_B, k_{t1}) \, .
\eeq
The quantities entering \eq{eq:master-kt-space} are as follows.
The Sudakov radiator $R(k_{t1})$ can be written as
\beq
\label{eq:sud2}
R(k_{t1})
\, = \,
- \, L \, g_1(\lambda) -
\sum_{n=0}^\infty
\Big(
\frac{\as}\pi
\Big)^n \, g_{n+2}(\lambda) \,,
\eeq
where $L = \ln(Q/k_{t1})$, $\lambda=\as \, \beta_0 \, L$, $\beta_0$ is the first
coefficient of the QCD beta function, and $Q \sim \mll$ is the resummation
scale. Variations of the latter can be used to gauge the impact of missing
higher-order logarithmic terms. For the predictions below, we choose
$Q= \mll/2$. In order to reach N$^k$LL logarithmic accuracy, all functions up
to $g_{k+1}(\lambda)$ need to be included. Expressions up to $g_4(\lambda)$ can
be found in appendix~B of~\cite{Bizon:2017rah}. The luminosity factor
${\cal L}(k_{t1})$ embeds the Born matrix element and the PDFs, as well as the
hard-virtual function $H$ and the collinear coefficient functions $C_{ab}$:
\beq
\label{eq:lumi}
{\cal L}(k_{t1})
\, = \,
\sum_{ijcd}
|{\cal M}_B|^2_{cd} \, \,
C_{ci}
\otimes
f_i(k_{t1})
\, \,
C_{d\hspace{0.08mm}j}
\otimes
f_j(k_{t1})
\, \,
H
\, .
\eeq
By including $H$ and $C_{ab}$ at ${\cal O}(\alpha_s^k)$, which formally
represents an N$^{k+1}$LL effect, the entire $\alpha_s^n L^{2(n-k)}$ logarithmic
tower is correctly described, which is referred to as N$^k$LL$'$ accuracy.
In this work we systematically employ \rad predictions with  N$^3$LL$'$
accuracy.
The last ingredient in \eq{eq:master-kt-space} is the function
${\cal F}(v, \Phi_B, k_{t1})$, which describes an arbitrary number of resolved
soft and/or collinear real emissions with a transverse momentum smaller than
$k_{t1}$, and whose inclusion is necessary starting from NLL accuracy. 
The full expression for this function, up to N$^3$LL$'$ in QCD, is presented
in~\cite{Re:2021con}, using ingredients extracted from \cite{Catani:2011kr,
  Catani:2012qa,Gehrmann:2014yya,Luebbert:2016itl,Echevarria:2016scs,
  Li:2016ctv,Vladimirov:2016dll,Moch:2017uml,Moch:2018wjh,Lee:2019zop,
  Luo:2019bmw,Henn:2019swt,Bruser:2019auj,Henn:2019rmi,Luo:2019szz,
  vonManteuffel:2020vjv,Ebert:2020yqt,Luo:2020epw}.

In order to smoothly switch off the resummation in the hard region
$k_{t1} \geq Q$, \rad{} replaces the resummed logarithm $L = \ln(Q/k_{t1})$
with a modified form $\tilde{L} = \ln[(Q/k_{t1})^p + 1]/p$ (with $p=6$ by
default),
accompanied by a Jacobian factor. This introduces controlled power corrections
that preserve the logarithmic accuracy and cancel exactly upon matching
to fixed order. The \rad code allows for several options for matching the
resummed calculation to fixed-order predictions. Here we use as default a
standard additive matching, performed at the differential level
(see~\cite{Re:2021con} for details):
\beq\label{eq:matching}
\frac{{\rm d}\sigma_{\rm RES+FO}}{{\rm d}\ptll}
\, = \,
\frac{{\rm d}\sigma_{\rm FO}}{{\rm d}\ptll}
\, + \, 
\frac{{\rm d}\sigma_{\rm RES}}{{\rm d}\ptll}
\, - \, 
\bigg[
\frac{{\rm d}\sigma_{\rm RES}}{{\rm d}\ptll}
\bigg]_{\rm FO}\, .
\eeq
In \eq{eq:matching} ${\rm d}\sigma_{\rm RES}$, ${\rm d}\sigma_{\rm FO}$,
and $[{\rm d}\sigma_{\rm RES}]_{\rm FO}$ denote the resummation contribution, the
fixed-order spectrum, and the perturbative expansion of the resummation,
respectively.

\begin{figure}[!t]
  \includegraphics[width=0.49\textwidth]{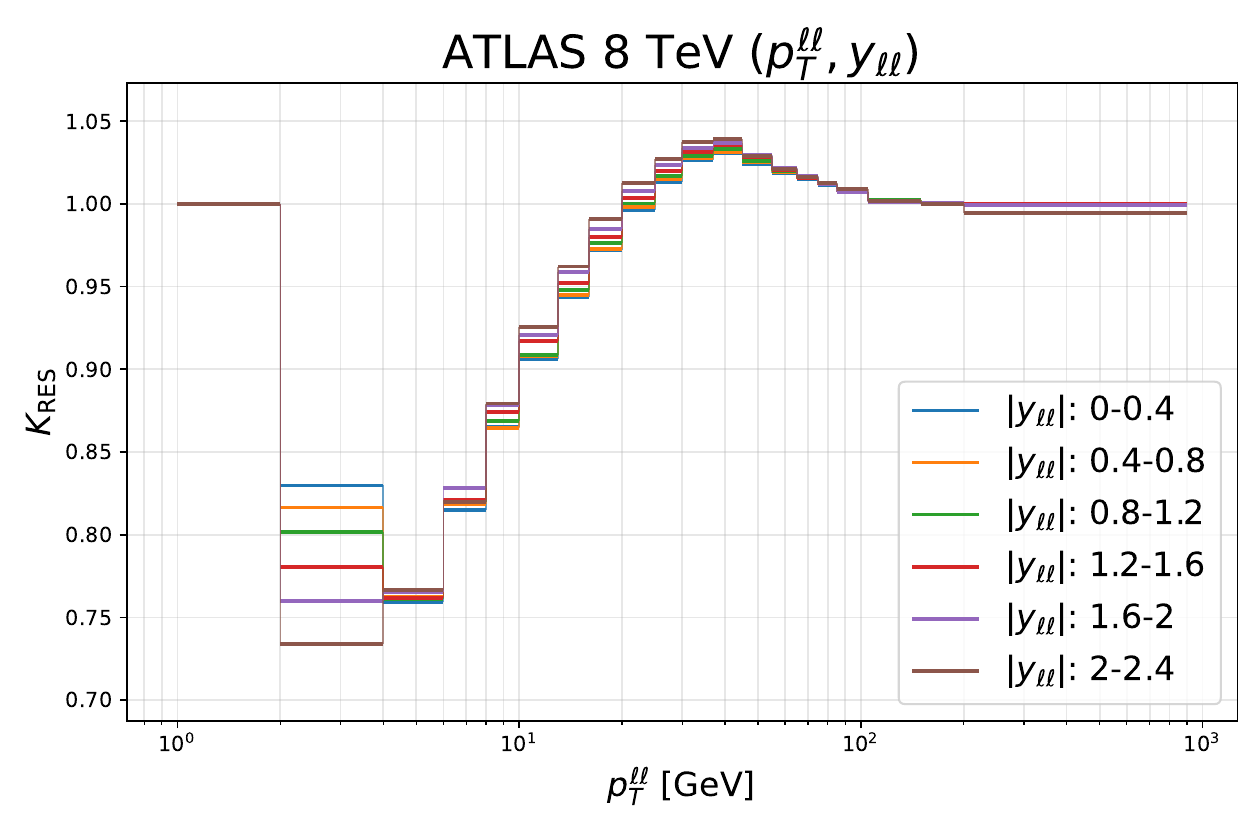}
  \includegraphics[width=0.49\textwidth]{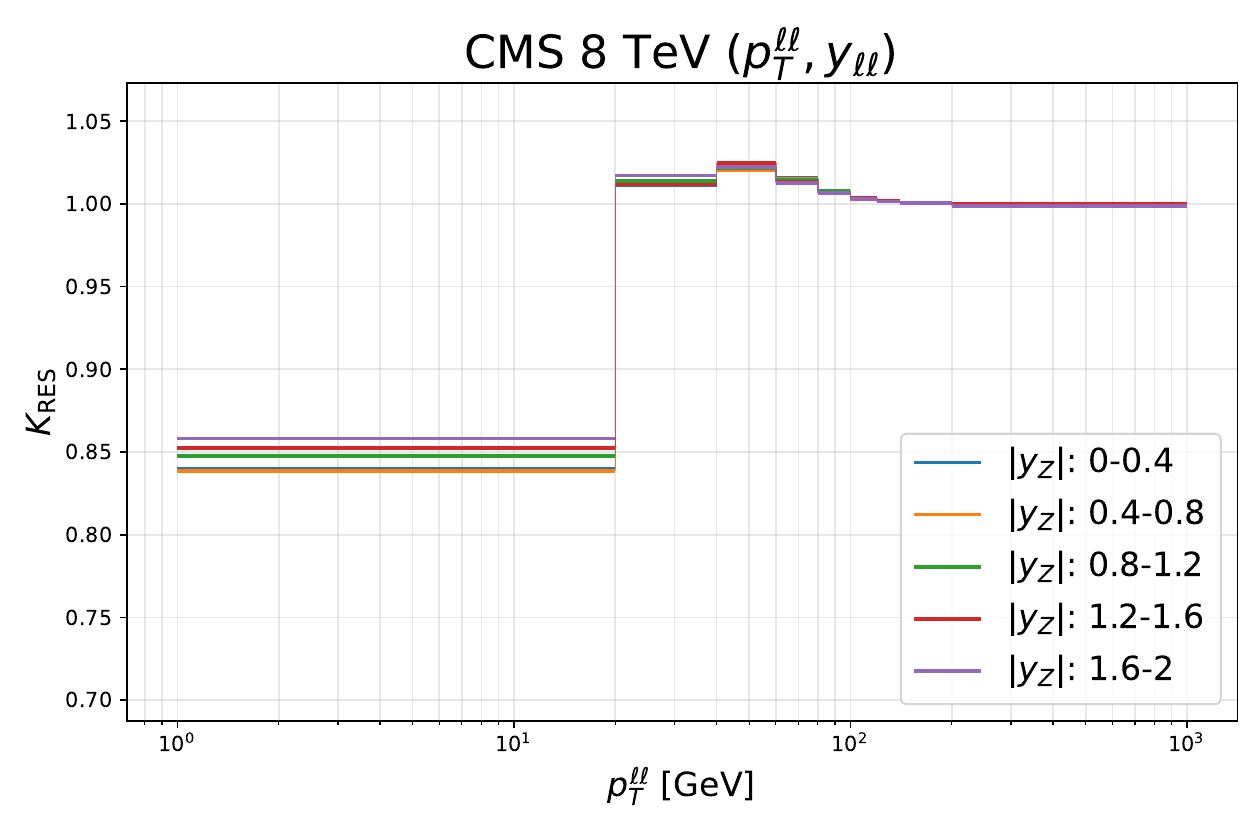}\\
  \includegraphics[width=0.49\textwidth]{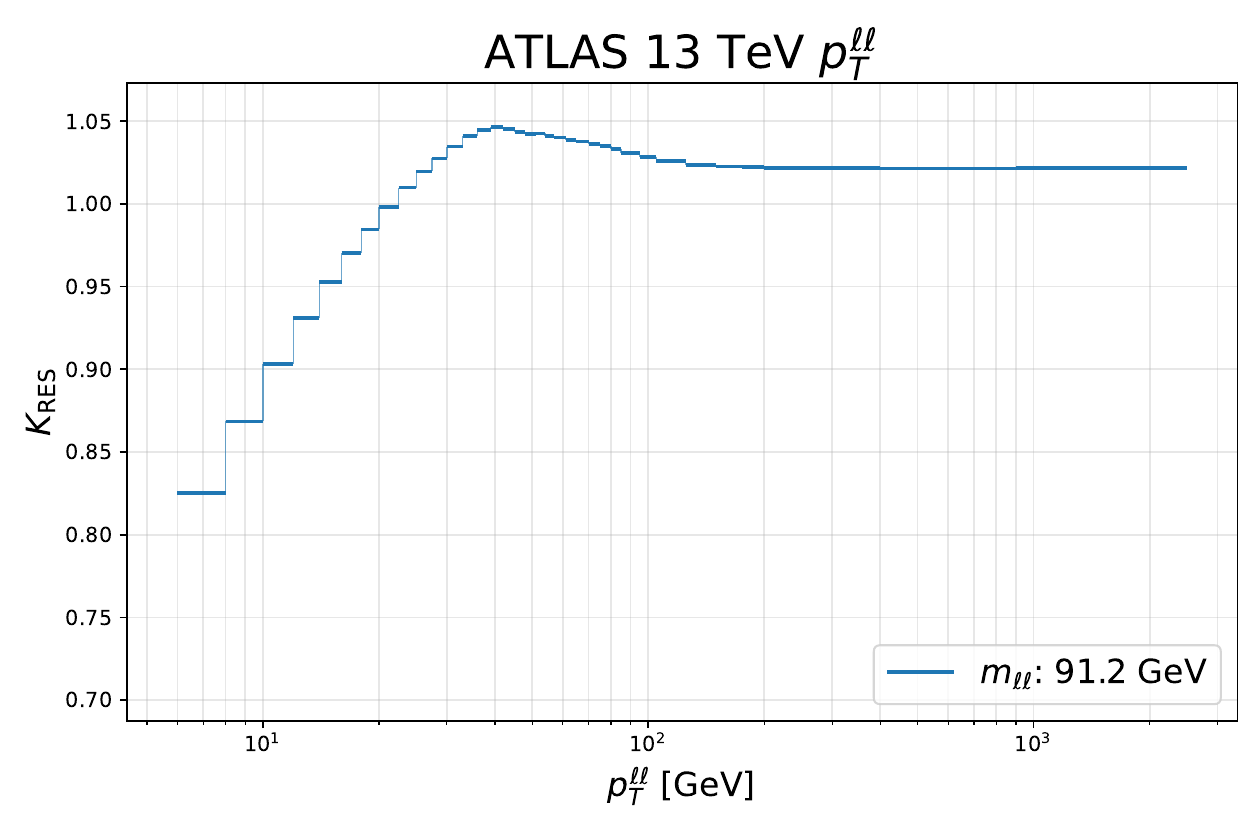}
  \includegraphics[width=0.49\textwidth]{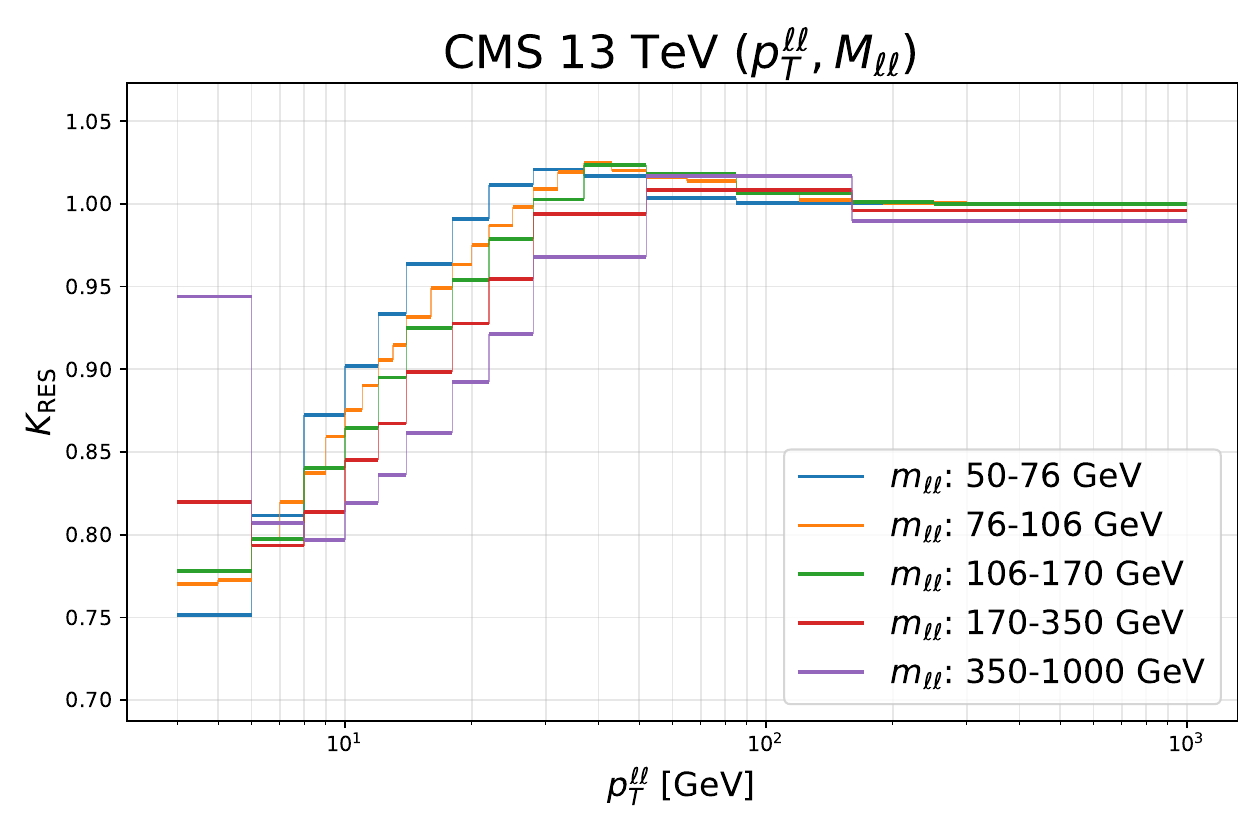}\\ 
  \caption{The resummation $K$-factor $K_{\rm RES}$ defined in \eq{eq:res_K-factor} for the
    $Z$-boson transverse momentum distributions measured by ATLAS and CMS at
    8 and 13~TeV, as a function of $\ptll$. PDFs are taken from the
    nominal NNLO set of~\cite{Ball:2025xgq}.}
  \label{fig:res_K-factor}
\end{figure}

Matched \rad predictions are included upon the fixed-order ones discussed in
Sect.~\ref{subsubsec:NNLOjet} by means of a resummation $K$-factor,
\begin{equation}
  K_{\rm RES}
  =
  \frac{d\sigma_{\rm RES+FO}/d\ptll}{d\sigma_{\rm FO}/d\ptll}\,,
  \label{eq:res_K-factor}
\end{equation}
where the fixed order is accurate to NNLO in the strong coupling, and
PDFs are taken from the nominal NNLO set of~\cite{Ball:2025xgq}.
In Fig.~\ref{fig:res_K-factor}, we display the $K$-factor in
\eq{eq:res_K-factor} as a function of $\ptll$ for all the rapidity bins
of the ATLAS and CMS 8 and 13~TeV measurements discussed in
Sect.~\ref{subsec:data}. The shape of $K_{\rm RES}$ is
roughly the same for ATLAS and CMS at 8 and 13~TeV. It starts with a
15-25\% depletion of the cross section at small $\ptll$, it then rises
to a 2-5\% enhancement around $\ptll\sim 40$~GeV, and it finally decreases
to one at large $\ptll$, roughly $\ptll\gtrsim 100$~GeV.
Interestingly, $K_{\rm RES}$ amounts to a few percent even in the region
which is not excluded by the nominal 30 GeV cut adopted in~\cite{Ball:2025xgq}.
We will therefore study its relevance in a PDF fit in Sect.~\ref{subsec:res}.

%-------------------------------------------------------------------------------
%-------------------------------------------------------------------------------

\subsubsection{MHOUs: theory covariance matrix}
\label{subsubsec:th_cov}

We supplement our fixed-order and resummed predictions with an estimate of
missing higher-order uncertainties. To this purpose, we use the theory
covariance matrix formalism developed in~\cite{NNPDF:2019vjt,NNPDF:2019ubu},
see also~\cite{NNPDF:2024dpb}. The idea is to construct a theoretical
covariance matrix from renormalisation- and factorisation-scale variations, and
to add it to the experimental covariance matrix. The sum of the two covariance
matrices is then used to sample pseudodata replicas and to compute the $\chi^2$
used for parameter optimisation. As in~\cite{NNPDF:2024dpb}, we compute the
elements of the theory covariance matrix using a 7-point prescription, in
which renormalisation and factorisation scales are varied by a factor 1/2 and
2 with respect to the central scale choice, see eqs.~(4-18)-(4.19)
in~\cite{NNPDF:2019ubu}. Renormalisation-scale variations govern the scale
dependence of matrix elements, while factorisation-scale variations govern the
scale dependence of DGLAP evolution equations. The former are correlated only
across data points that belong to the same physical processes while the latter
are correlated across all data points.

We note that, when supplementing fixed-order computations with resummation
corrections, physical observables depend on an additional scale, the resummation
scale introduced below \eq{eq:sud2}. In principle, a consistent account of
missing higher-order uncertainties would require to incorporate variations of
the latter scale in the construction of the theory covariance matrix, along
with factorisation and renormalisation scales. In this work, we neglect this
further variation, on the basis that, as we will show in Sect.~\ref{subsec:res},
its effect is marginal. We also remark that, when constructing the theory
covariance matrix, we use resummation $K$-factors that are matched to
scaled-varied fixed-order predictions. In this way, resummation $K$-factors
depend on renormalisation- and factorisation-scale variations.

\subsection{Methodological framework}
\label{subsec:NNPDF}

The methodological framework used for our PDF fits is based on that presented
in~\cite{NNPDF:2024dpb}. Its salient points, which define the so-called
NNPDF4.0 methodology, are summarised as follows. Parton distributions are
parametrised in terms of a single feed-forward deep neural network with eight
independent outputs. These correspond to the gluon PDF, and to singlet and
non-singlet PDF combinations, including the contribution of a non-perturbative
component of the charm quark. The parametrisation scale is taken slightly above
the charm quark mass, namely $Q_0=1.65$~GeV. The neural network is optimised
to the data by means of adaptive gradient descent. Hyperparameters controlling
the neural network architecture and the optimisation algorithm are determined
from a hyperparameter scan on $K$-folds. The set of hyperparameters used in this
work is the same as that discussed in Sect.~3.2 of~\cite{NNPDF:2021njg}.
The methodology also incorporates theoretical constraints, namely momentum and
valence sum rules, and positivity and integrability constraints. The former are
implemented by construction in the PDF parametrisation, while the latter are
enforced by regularising the loss function with Lagrange multipliers. The
complete set of theoretical constraints is the same as in Sects.~3.1.2, 3.1.3,
and 3.1.4 of~\cite{NNPDF:2021njg}. Experimental and theoretical uncertainties
are propagated to PDFs by generating an ensemble of pseudodata replicas,
sampled from a multi-Gaussian distribution with central value equal to the
central value of the experimental data and covariance equal to the sum of
experimental and theoretical covariance matrices. PDF replicas are optimised to
each pseudodata replica in such a way that a final ensemble of PDFs is
produced. Expectation values and uncertainties on physical observables can then
be computed as mean values and standard deviations over the ensemble of PDF
replicas. All the fits presented in this work correspond to ensembles of
100 Monte Carlo replicas. The internal consistency of the methodology
is finally validated by means of closure tests~\cite{DelDebbio:2021whr},
whereas its generalisation to unseen data sets is checked by means of future
tests~\cite{Cruz-Martinez:2021rgy}.

The framework presented in~\cite{NNPDF:2024dpb} otherwise differs from the
original NNPDF4.0 methodology of~\cite{NNPDF:2021njg} in a few respects.
Specifically, it features a new format for handling data sets, which led to
the identification and correction of a small number of minor issues in the
data implementation, as well as a new pipeline for the computation of theory
predictions~\cite{Barontini:2023vmr}. The latter includes, in particular, a
new treatment of heavy-quark mass effects, which differs from the
previous implementation only by subleading terms~\cite{Barontini:2024xgu},
and employs the exact solution of the QCD renormalisation-group equation.
The impact of these changes was assessed in Appendix A of
Ref.~\cite{NNPDF:2024djq},
and was found to be very limited. As we mentioned above, we also update
the values of the  physical parameters, standardise renormalisation and
factorisation scale choices in DY data, drop $K$-factors in favour of
exact computations to include NNLO QCD corrections, and move to the
full-colour computation for dijet measurements. The impact of these
changes will be discussed in Sect.~\ref{subsec:baseline}.

%% file: sec-results.tex
\section{Results}
\label{sec:results}

In this section, we present the results of a series of PDF fits based on the
experimental, theoretical and methodological input discussed in
Sect.~\ref{sec:input}. We specifically proceed with a staged series of
fits, through which we assess the impact of three classes of different,
yet complementary, effects.

%-------------------------------------------------------------------------------
\begin{table}[!t]
  \centering
  \footnotesize
  \renewcommand{\arraystretch}{1.2}
  \begin{tabularx}{\textwidth}{lXc}
    \toprule
    ID & Description & Sect. \\
    \midrule
    4.0
    &
    Reference NNPDF4.0 fit, corresponding to the default fit
    of~\cite{Ball:2025xgq}.
    &
    \ref{subsec:baseline}
    \\
    4.0no1pc
    &
    A fit in which the additional uncorrelated 1\% uncertainty on ATLAS and
    CMS 8~TeV $Z$-boson transverse-momentum measurements is removed, and which
    is otherwise equivalent to the 4.0 fit.
    &
    \ref{subsec:baseline}
    \\
    4.0NNLOjet
    &
    A fit in which $K$-factors are dismissed in favour of exact and
    numerically stable computations for NNLO QCD corrections; the central
    factorisation and renormalisation scales are set to the transverse mass
    for all DY measurements; full-colour computations are used for dijet
    measurements; $N$-jettiness computations are replaced by
    antenna-subtraction \nnlojet computations for $Z$-boson
    transverse-momentum distributions. Otherwise, this fit is equivalent
    to the 4.0 fit.
    &
    \ref{subsec:baseline}
    \\
    baseline
    &
    A fit in which the additional uncorrelated 1\% uncertainty on ATLAS and
    CMS 8~TeV $Z$-boson transverse-momentum measurements is removed, and which
    is otherwise equivalent to the 4.0NNLOjet fit.
    &
    \ref{subsec:baseline}
    \\
    \midrule
    13AT
    &
    A fit in which the 13~TeV ATLAS $Z$-boson transverse-momentum
    measurements~\cite{ATLAS:2019zci} are added to the data set of the baseline
    fit, to which it is otherwise equivalent.
    &
    \ref{subsec:13TeV}
    \\
    13CM
    &
    A fit in which the 13~TeV CMS $Z$-boson transverse-momentum
    measurements~\cite{CMS:2022ubq} are added to the data set of the baseline
    fit, to which it is otherwise equivalent.
    &
    \ref{subsec:13TeV}
    \\
    13ATCM
    &
    A fit in which the 13~TeV ATLAS and CMS $Z$-boson transverse-momentum
    measurements~\cite{ATLAS:2019zci,CMS:2022ubq} are added to the data set of
    the baseline fit, to which it is otherwise equivalent.
    &
    \ref{subsec:13TeV}
    \\
    \midrule
    RES
    &
    A fit in which the small-$\ptll$ resummation $K$-factors
    of~\eq{eq:res_K-factor} are added to the 8~TeV ATLAS and CMS $Z$-boson
    transverse-momentum measurements, and
    which is otherwise equivalent to the baseline fit. As in all other fits, we
    require $\ptll\geq\ptmll$, with $\ptmll=30$~GeV.
    &
    \ref{subsec:res}
    \\
    13RES
    &
    A fit in which the small-$\ptll$ resummation $K$-factors
    of~\eq{eq:res_K-factor} are added to the 8 and 13~TeV ATLAS and CMS
    $Z$-boson transverse-momentum measurements,
    and which is otherwise equivalent to the 13ATCM fit. We require
    $\ptll\geq\ptmll$, with $\ptmll=30$~GeV.
    &
    \ref{subsec:res}
    \\
    13RESAT, 13RESCM
    &
    Two fits in which the small-$\ptll$ resummation $K$-factors
    of~\eq{eq:res_K-factor} are added to the 8 and 13~TeV ATLAS or CMS
    $Z$-boson transverse-momentum measurements, included
    one at a time, and which are otherwise equivalent to the 13AT and 13CM fits,
    respectively. We require $\ptll\geq\ptmll$, with $\ptmll=30$~GeV.
    &
    \ref{subsec:res}
    \\
    RES20, RES10, RES4
    &
    Fits in which the cut on $\ptll\geq\ptmll$ for the
    $Z$-boson transverse momentum is set to $\ptmll=20$, 10,
    and 4~GeV, and which are otherwise equivalent to the RES fit.
    &
    \ref{subsec:res}
    \\
    13RES20, 13RES10, 13RES4
    &
    Fits in which the cut on $\ptll\geq\ptmll$ for the
    $Z$-boson transverse momentum is set to $\ptmll=20$, 10,
    and 4~GeV, and which are otherwise equivalent to the 13RES fit.
    &
    \ref{subsec:res}
    \\
    \bottomrule
  \end{tabularx}
  \caption{A summary of the fits performed in this work. For each of them, we
    indicate a shorthand ID, provide a brief description, and report the section
    in which they are discussed.}
  \label{tab:fits}
\end{table}
%-------------------------------------------------------------------------------

The first class of effects is related to the changes in the numerical
treatment of the experimental data already contained in our reference
fit~\cite{Ball:2025xgq}.
As previously mentioned, such changes include: an update of the values of
the physical parameters; the consistent adoption of the transverse mass
as the central renormalisation and factorisation scale for all DY data sets;
the use of full-colour (in lieu of the leading-colour) predictions for dijet
measurements; and the replacement of NNLO QCD $K$-factors with exact,
numerically stable NNLO calculations.
On top of these changes, that affect the entire data set
in~\cite{Ball:2025xgq}, the 8~TeV ATLAS and CMS $Z$-boson
transverse-momentum distributions~\cite{ATLAS:2015iiu,CMS:2015hyl}
are specifically concerned with the replacement of the $N$-jettiness
prediction by the \nnlojet antenna-subtraction computation, and with the
possibility of removing the $1\%$ uncertainty that was previously introduced
to account for the numerical instabilities of the former.
We will therefore perform a series of fits in which the effect of
these changes is assessed incrementally.
The fit that cumulatively incorporates them all represents our baseline fit,
against which we will compare all subsequent ones.

The second class of effects is related to the impact of the new 13~TeV
ATLAS and CMS $Z$-boson transverse-momentum
distributions~\cite{ATLAS:2019zci,CMS:2022ubq}. We will perform two fits,
each including one of the new data sets, and a third fit in which both data sets
are added simultaneously to the baseline fit obtained previously. By comparing
these fits, we will assess the impact of each data set on the PDFs, as well as
their compatibility with the rest of the data and with each other.

The third class of effects is related to the impact on the fit quality
and on PDFs of small-$\ptll$ resummation corrections to the 8
and 13~TeV ATLAS and CMS $Z$-boson transverse-momentum distributions.
We will first repeat the baseline fit with the addition of resummation
corrections only, and then perform a set of fits including both resummation
corrections and the 13~TeV measurements. We consider the case in which the
13~TeV ATLAS and CMS measurements are added separately or together.
In the cases in which both the 8~TeV ATLAS and CMS, or both the 8 and 13~TeV
ATLAS and CMS measurements are included, we will repeat the fits by varying the
kinematic cut on $\ptll$ that affects
$Z$-boson transverse-momentum measurements, reducing it incrementally from
the nominal 30~GeV down to 20, 10 and 4~GeV. We will then investigate the impact
of resummation corrections on the PDFs as the data set and the applied kinematic
cuts are varied.

To facilitate the following discussion, all fits are summarised in
Table~\ref{tab:fits}. For each fit, we provide a shorthand identifier, a brief
description, and the section in which it is discussed. We next proceed to
examine each class of the aforementioned effects in turn.

\subsection{Baseline treatment of theoretical predictions}
\label{subsec:baseline}

\begin{figure}[!p]
  \centering
  \includegraphics[width=0.48\textwidth]{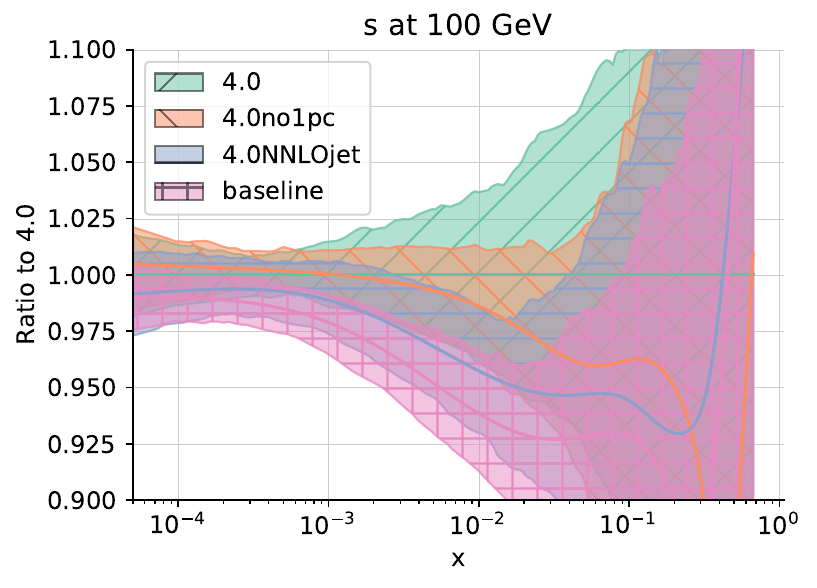}
  \includegraphics[width=0.48\textwidth]{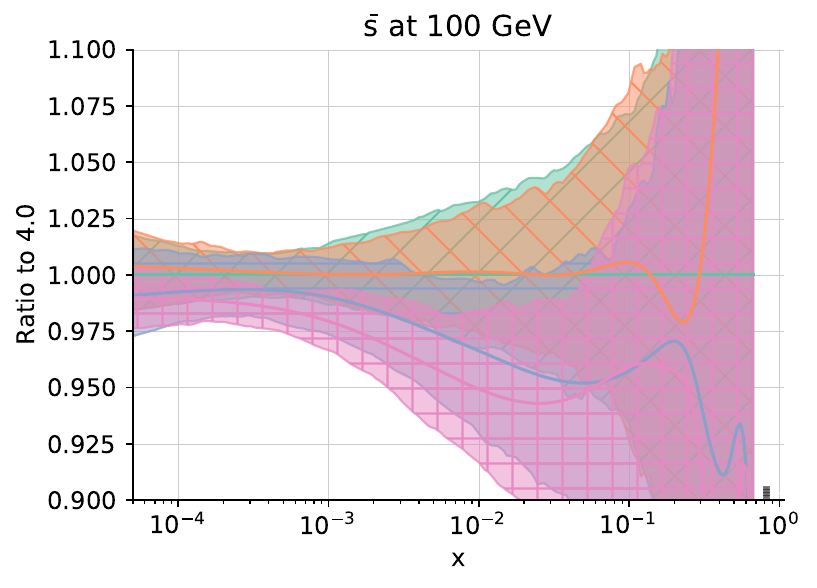}\\
  \includegraphics[width=0.48\textwidth]{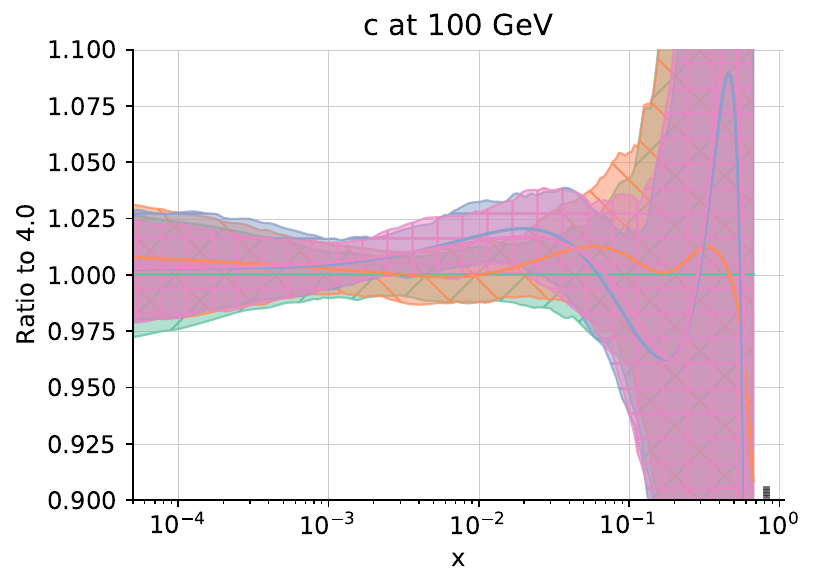}
  \includegraphics[width=0.48\textwidth]{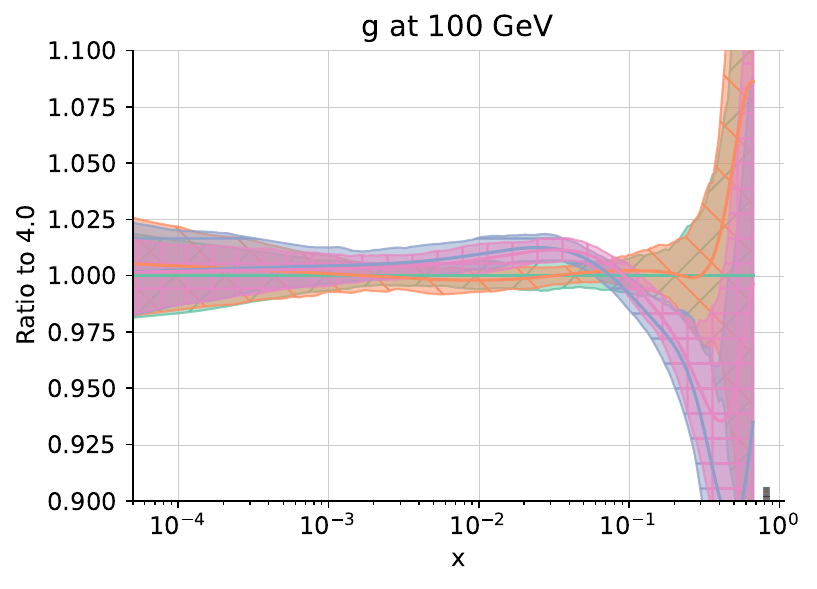}\\
  \caption{Comparison of the strange quark and anti-quark, charm quark and
    gluon PDFs obtained from the fits listed in the top four rows of
    Table~\ref{tab:fits}.  Parton distributions are displayed at an energy
    scale of $Q=100$~GeV as a function of the proton momentum fraction $x$,
    and are normalised to the 4.0 fit. Error bands correspond to 68\%
    confidence levels.}
  \label{fig:PDFs_8TeV_baseline}
\end{figure}

We start by discussing the first class of effects outlined above, which
collectively lead to the definition of our baseline fit. To this purpose, we
analyse the fits discussed in the top four rows of Table~\ref{tab:fits}.
In Table~\ref{tab:chi2_8TeV_baseline}, we display the values of the $\chi^2$
per number of data points, $N_{\rm dat}$, for each of them. Data sets are
grouped by process (for their description, see for instance Sect.~3.1
of~\cite{NNPDF:2024dpb}), however we also single out the $Z$-boson
transverse-momentum measurements, which enter the DY NC category.
In Fig.~\ref{fig:PDFs_8TeV_baseline}, we compare the strange quark and
anti-quark, charm quark, and gluon PDFs obtained from the four fits of
Table~\ref{tab:chi2_8TeV_baseline}. We do not show PDFs of other partons, the
reason being that, in this case, differences among fits are essentially
immaterial. Parton distributions are displayed at an energy scale 
$Q=100$~GeV as a function of the proton momentum fraction $x$, and are
normalised to the 4.0 fit. Error bands correspond to the 68\% confidence level.
Inspection of Table~\ref{tab:chi2_8TeV_baseline} and
Fig.~\ref{fig:PDFs_8TeV_baseline} leads us to some observations, that we
summarise as follows. 

%-------------------------------------------------------------------------------
\begin{table}[!p]
  \centering
  \footnotesize
  \renewcommand{\arraystretch}{1.2}
  \begin{tabularx}{\textwidth}{Xrccccc}
    Data set
    & $N_{\rm dat}$
    & \rotatebox{90}{4.0}
    & \rotatebox{90}{4.0no1pc}
    & \rotatebox{90}{4.0NNLOjet}
    & \rotatebox{90}{baseline}\\
    \toprule
    DIS NC
    & 2100 & 1.19 & 1.19 & 1.19 & 1.19 \\
    DIS CC
    &  989 & 0.89 & 0.90 & 0.89 & 0.89 \\
    DY NC
    &  671 & 1.15 & 1.35 & 1.05 & 1.23 \\
    \quad ATLAS $\ptll$ 8~TeV ($\ptll,\yll$)
    &   44 & 0.92 & 1.06 & 1.08 & 1.24 \\
    \quad ATLAS $\ptll$ 8~TeV ($\ptll,\mll$)
    &   48 & 0.69 & 1.97 & 0.65 & 1.97 \\
    \quad CMS $\ptll$ 8~TeV ($\ptll,\yll$)
    &   28 & 1.42 & 3.63 & 0.87 & 1.79 \\
    DY CC
    &  157 & 1.34 & 1.36 & 1.06 & 1.07 \\
    top-quark pair
    &   64 & 1.54 & 1.36 & 1.00 & 1.00 \\
    single-inclusive jets
    &  356 & 0.80 & 0.82 & 0.88 & 0.87 \\
    dijets
    &  144 & 1.65 & 1.65 & 1.54 & 1.52 \\
    photon
    &   53 & 0.67 & 0.66 & 0.84 & 0.82 \\
    single top-quark
    &   17 & 0.38 & 0.38 & 0.17 & 0.17 \\
    \midrule
    Total
    & 4551 & 1.12 & 1.16 & 1.10 & 1.12 \\
    \bottomrule
  \end{tabularx}
  \caption{The values of the $\csq$ per number of data points $N_{\rm dat}$
    for each of the fits in the top four rows of Table~\ref{tab:fits}.
    Data sets are aggregated by process, however we single out the $Z$-boson
    transverse-momentum measurements that enter the DY NC category.}
  \label{tab:chi2_8TeV_baseline}
\end{table}
%-------------------------------------------------------------------------------

%-------------------------------------------------------------------------------
%-------------------------------------------------------------------------------

We first compare the 4.0 and 4.0NNLOjet fits. They rely on the same data set,
but differ for the details of the theoretical computations summarised in
Table~\ref{tab:fits}. Overall, the data set is more consistently described
in the 4.0NNLOjet fit than in the 4.0 fit; indeed the value of the total $\csq$
per number of data points decreases from 1.12 to 1.10. For $N_{\rm dat}=4551$
data points, this improvement corresponds to about one standard deviation of
the $\csq$ distribution. The improvement is mainly driven by an improvement of
both NC (including $Z$-boson transverse momentum) and CC DY, top-quark pair,
and dijet data sets. This is not surprising: the use of a standardised
renormalisation and factorisation scale for DY data, of the more accurate 
antenna-subtraction computation for $Z$-boson transverse-momentum data, and of
the full-colour computation for dijet data are all ingredients that improve
the accuracy of the theoretical framework and, as a consequence, of the overall
consistency of the data set. However, we note that the improvement in fit
quality comes with a shift of the PDF central value: for the strange quark and
anti-quark PDFs this amounts to a depletion up to about one sigma in units of
the PDF uncertainty around $x\sim 0.03$; for the charm quark and for the gluon
PDFs, this amounts to an increase up to 1.5 sigma around a similar value
of $x$.
We have traced these effects, respectively, to the changes in the
theoretical treatment of DY and dijet data. A more detailed discussion will
be part of the forthcoming NNPDF4.1 release~\cite{Ball2026:NNPDF41}.

We then compare two pairs of fits: 4.0 versus 4.0no1pc, and 4.0NNLOjet versus
the baseline fit. The two fits in each pair differ only by the inclusion or
omission of the 1\% uncertainty introduced to account for potential numerical
instabilities in the theoretical predictions for the $Z$-boson
transverse-momentum measurements.
As expected, lifting this uncertainty leads to a deterioration of the global
$\csq$, which is essentially driven by a deterioration of the $\csq$ of the
$Z$-boson transverse-momentum data.
The size of this deterioration, however, depends on the underlying computation:
it is larger when using the $N$-jettiness computation (the value of  the global
$\csq$ increases from 1.12 to 1.16), whereas it is smaller when using
the \nnlojet antenna-subtraction computation (the value of the global
$\csq$ increases from 1.10 to 1.12). We interpret this result as a confirmation
of the fact that, in our case of interest, the latter is numerically more
reliable than the former. Interestingly, the global fit quality of the baseline
fit is the same as the starting 4.0 fit. We finally note that lifting the
additional 1\% uncertainty on $Z$-boson transverse-momentum data
has a limited impact on PDFs, the only effect being a mild reduction of PDF
uncertainties for strange quarks and anti-quarks. This behaviour is independent
of whether the $N$-jettiness or the \nnlojet local-subtraction computation
is used. In light of all these considerations, we will use the settings of the
baseline fit as default for the data set and resummation studies that come next.

\subsection{Impact of 13~TeV $Z$-boson transverse-momentum data}
\label{subsec:13TeV}

\begin{figure}[!p]
  \centering
  \includegraphics[width=0.48\textwidth]{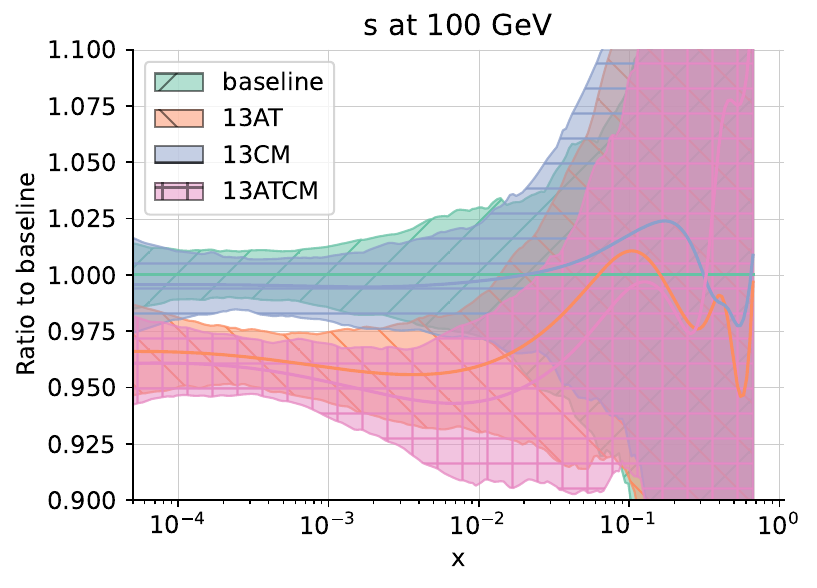}
  \includegraphics[width=0.48\textwidth]{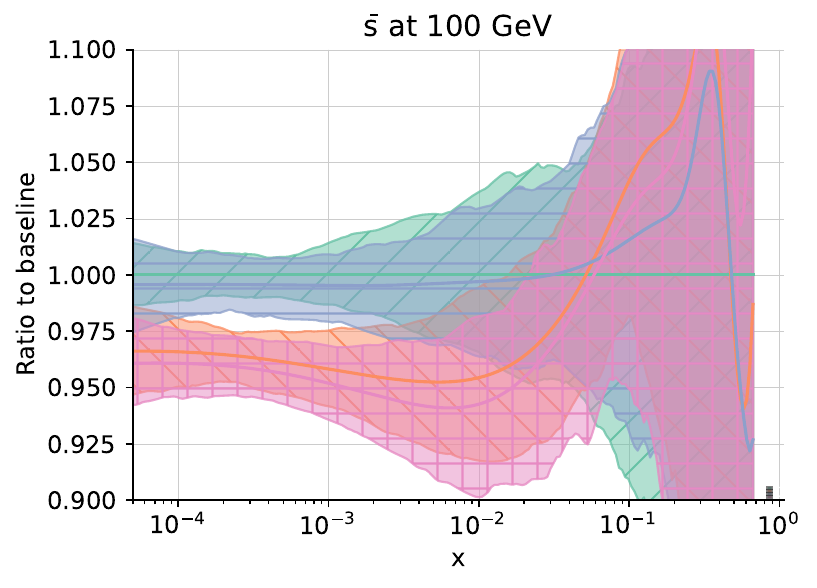}\\
  \includegraphics[width=0.48\textwidth]{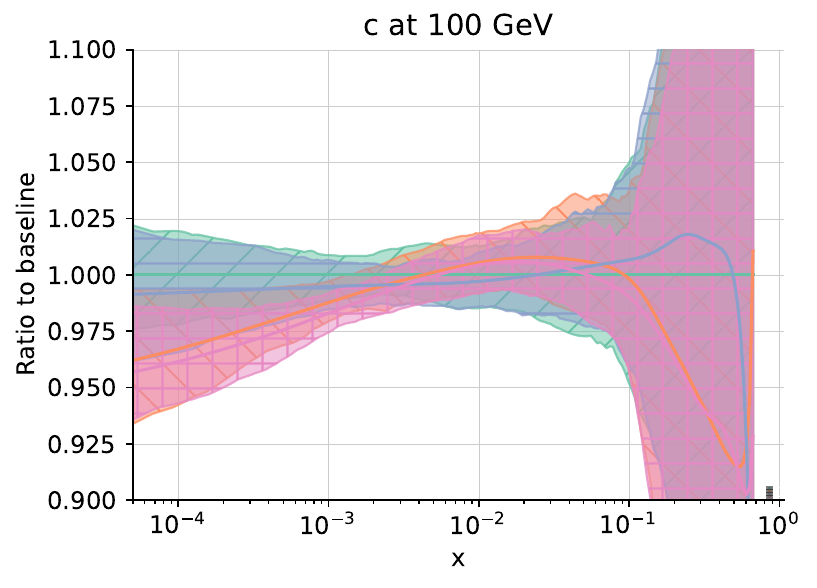}
  \includegraphics[width=0.48\textwidth]{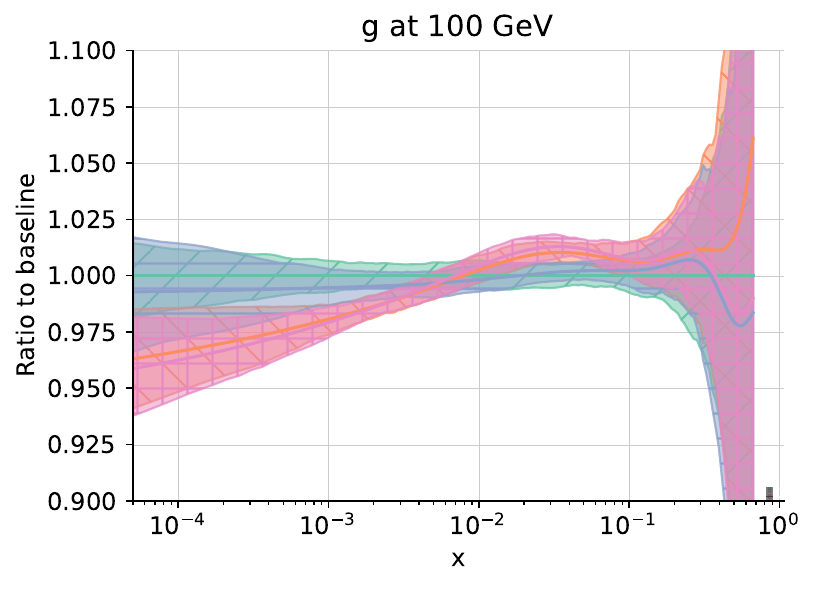}\\
  \caption{Comparison of the strange quark and anti-quark, charm quark and
    gluon PDFs obtained from the baseline fit and from the fits listed in the
    central three rows of Table~\ref{tab:fits}. The format is
    as in Fig.~\ref{fig:PDFs_8TeV_baseline}.}
  \label{fig:PDFs_8TeV_baseline+13TeV_analyses}
\end{figure}

We then discuss the second class of effects outlined above, that lead to the
assessment of the impact of the new $Z$-boson transverse-momentum data sets.
To this purpose, we analyse the three fits discussed in the central rows of
Table~\ref{tab:fits}, and compare them to the baseline fit.
In Table~\ref{tab:chi2_8TeV_baseline+13TeV_analyses}, we display the values of
the $\csq$ per number of data points, $N_{\rm dat}$, for each of them, whereas
in Fig.~\ref{fig:PDFs_8TeV_baseline+13TeV_analyses} we compare the corresponding
strange quark and anti-quark, charm quark, and gluon PDFs. The format of
Table~\ref{tab:chi2_8TeV_baseline+13TeV_analyses} and of
Fig.~\ref{fig:PDFs_8TeV_baseline+13TeV_analyses} is the same as
Table~\ref{tab:chi2_8TeV_baseline} and Fig.~\ref{fig:PDFs_8TeV_baseline},
respectively. Inspection of Table~\ref{tab:chi2_8TeV_baseline+13TeV_analyses}
and Fig.~\ref{fig:PDFs_8TeV_baseline+13TeV_analyses} leads us to the following
conclusions.

%-------------------------------------------------------------------------------
\begin{table}[!p]
  \centering
  \footnotesize
  \renewcommand{\arraystretch}{1.2}
  \begin{tabularx}{\textwidth}{Xrccccc}
    Data set
    & $N_{\rm dat}$
    & \rotatebox{90}{baseline}
    & \rotatebox{90}{13AT}
    & \rotatebox{90}{13CM}
    & \rotatebox{90}{13ATCM}\\
    \toprule
    DIS NC
    & 2100 & 1.19 & 1.21 & 1.20 & 1.20 \\
    DIS CC
    &  989 & 0.89 & 0.89 & 0.89 & 0.89 \\
    DY NC
    &  671 & 1.23 & 1.49 & 1.23 & 1.44 \\
    \quad ATLAS $\ptll$ 8~TeV ($\ptll,\yll$)
    &   44 & 1.24 & 1.17 & 1.22 & 1.17 \\
    \quad ATLAS $\ptll$ 8~TeV ($\ptll,\mll$)
    &   48 & 1.97 & 1.86 & 1.92 & 1.87 \\
    \quad CMS $\ptll$ 8~TeV ($\ptll,\yll$)
    &   28 & 1.79 & 1.84 & 1.81 & 1.85 \\
    \quad ATLAS $\ptll$ 13~TeV
    &   19 & ---  & 8.22 & ---  & 7.83 \\
    \quad CMS $\ptll$ 13~TeV
    &   16 & ---  & ---  & 1.15 & 0.89 \\
    DY CC
    &  157 & 1.07 & 1.06 & 1.07 & 1.05 \\
    top-quark pair
    &   64 & 1.00 & 0.94 & 0.98 & 0.91 \\
    single-inclusive jets
    &  356 & 0.87 & 0.89 & 0.88 & 0.89 \\
    dijets
    &  144 & 1.52 & 1.50 & 1.49 & 1.51 \\
    photon
    &   53 & 0.82 & 0.76 & 0.81 & 0.76 \\
    single top-quark
    &   17 & 0.17 & 0.15 & 0.17 & 0.15 \\
    \midrule
    Total
    & 4551 & 1.12 & 1.17 & 1.12 & 1.16 \\
    \bottomrule
  \end{tabularx}
  \caption{The values of the $\csq$ per number of data points $N_{\rm dat}$
    for the baseline fit and for each of the fits in the central three rows of
    Table~\ref{tab:fits}. The format is the same as that of
    Table~\ref{tab:chi2_8TeV_baseline}. Note that the total number of data
    points does not include the data points in the 13~TeV ATLAS and CMS
    $Z$-boson transverse-momentum data sets.}
  \label{tab:chi2_8TeV_baseline+13TeV_analyses}
\end{table}
%-------------------------------------------------------------------------------

%-------------------------------------------------------------------------------
%-------------------------------------------------------------------------------

We first compare the baseline and 13AT fits. We immediately notice that the
new ATLAS 13~TeV $Z$-boson transverse-momentum measurement is very badly
described, with a $\csq$ per number of data points of 8.22.\footnote{This
finding contradicts the result reported in~\cite{Chiefa:2025loi}, where the
description of the ATLAS 13~TeV $Z$-boson transverse-momentum measurement
was deemed to be optimal. Upon some investigations, we discovered that the
result of~\cite{Chiefa:2025loi} is possibly incorrect because of a mismatch
in the accounting of experimental uncertainties.} This fact
automatically degrades the global $\csq$ of the fit, which increases from
1.12 to 1.17, that is, by about 2.5 standard deviations of the $\csq$
distribution. The description of the other data sets in some cases
deteriorates slightly (for the DIS NC, single-inclusive jet, and 8~TeV CMS
$Z$-boson transverse-momentum measurements), whereas in some other cases it
improves (for top-quark pair, dijet, and the 8~TeV ATLAS $Z$-boson
transverse-momentum measurements). 

We have explicitly checked that the anomalously large $\csq$ of the 13~TeV
ATLAS $Z$-boson transverse-momentum measurement cannot be ascribed to an
ill-conditioned experimental covariance matrix (ensuing specifically from poor
control on uncertainty correlations): we have recomputed the $\csq$ with the
regularisation procedure outlined in~\cite{Kassabov:2022pps}, and found that
the $\csq$ remains essentially unchanged. We also remark that
EW corrections are unlikely to help. Indeed, on the one hand, we apply a
rather conservative cut, which excludes the large-$\ptll$ region where
these could possibly have an impact. On the other hand, even if we had included
them, these would have possibly brought the central value of the predictions
farther away from the experimental data, see Fig.~6 in~\cite{ATLAS:2019zci}.

An additional investigation of the internal consistency of the 13~TeV ATLAS
$Z$-boson transverse-momentum measurement and of its compatibility with the
rest of the data set in the baseline fit is presented in
Appendix~\ref{app:weighted_fit}. For now, inspecting
Fig.~\ref{fig:PDFs_8TeV_baseline+13TeV_analyses}
we note that the inclusion of this data set distorts the PDFs in a significant
way with respect
to the baseline fit. The central value of the strange quark and anti-quark
PDFs is depleted by about 5\% for $x\lesssim 0.01$, in a region where the
nominal PDF uncertainty is about half of that. The effect on the charm quark PDF
is less significant, as it is always encompassed by the PDF uncertainty,
whereas the impact on the gluon PDF is more concerning. In this case, the shape
of the PDF changes significantly, with a suppression of about 2.5\% for
$x\lesssim 0.001$, and an enhancement of 1\% for $x\sim 0.03$. The latter
effect may be particularly delicate, because it will directly affect predictions
for the gluon-fusion Higgs cross section.

We then compare the baseline and 13CM fit. We observe that the new CMS 13~TeV
$Z$-boson transverse-momentum measurement is fairly well described, with a
$\csq$ per number of data points of order one. The overall fit quality is the
same as that of the baseline fit. The impact on PDFs is very mild, with at
most a very marginal reduction of the strange quark, anti-quark, and gluon
PDF uncertainties.

We finally compare the baseline and the 13ATCM fits. As expected, this
comparison closely resembles that between the baseline and the 13AT fit.
Being the 13~TeV ATLAS $Z$-boson transverse-momentum measurement much more
precise than its CMS counterpart, the fit that includes both data sets is
essentially the same as the fit that includes only the ATLAS data set.
Therefore, the same considerations laid out above also apply to this case.
Whether the inclusion of small-$\ptll$ resummation corrections improves
this state of affairs will be discussed next.

\subsection{Impact of resummation corrections}
\label{subsec:res}

We finally focus on the third class of effects outlined above, namely the impact
of small-$\ptll$ resummation corrections on $Z$-boson transverse-momentum
data sets. We separate two cases. In the first case, we study the impact of
resummation corrections as the $Z$-boson transverse-momentum data sets are
varied, but the cut on $\ptll$ is kept fixed to
$\ptll\geq 30$~GeV. In the second case, we study the impact of
resummation corrections keeping the data set fixed (either to only 8~TeV
or to both 8~TeV and 13~TeV data) as the $\ptll$ cut is varied.

\subsubsection{Fixed cuts}
\label{subsubsec:fixed_cut}

\begin{figure}[!p]
  \centering
  \includegraphics[width=0.48\textwidth]{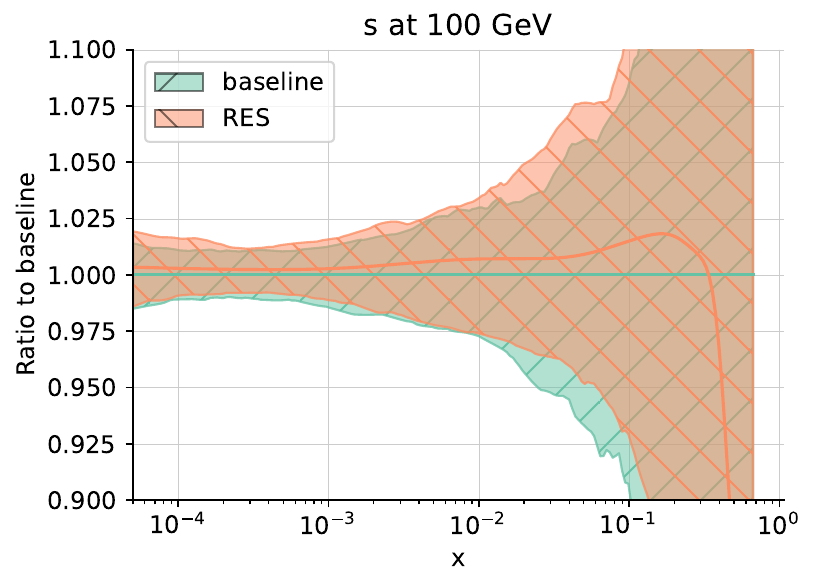}
  \includegraphics[width=0.48\textwidth]{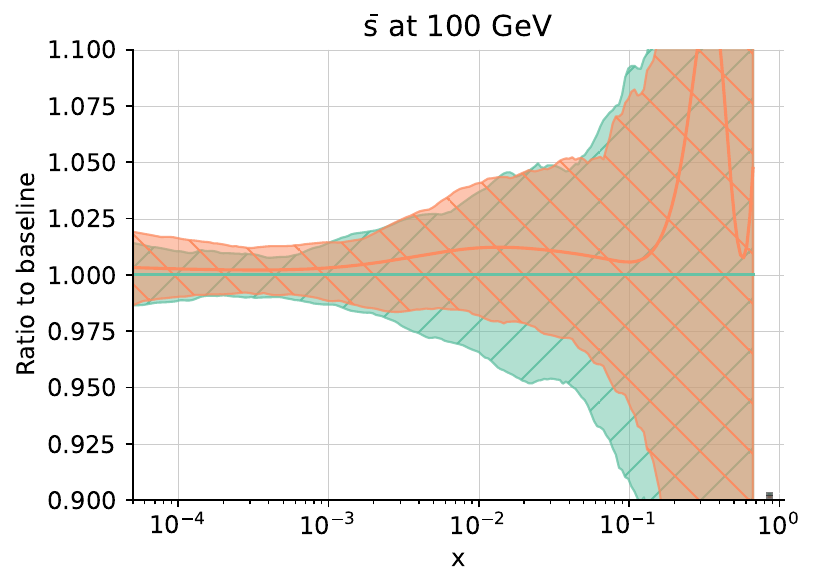}\\
  \includegraphics[width=0.48\textwidth]{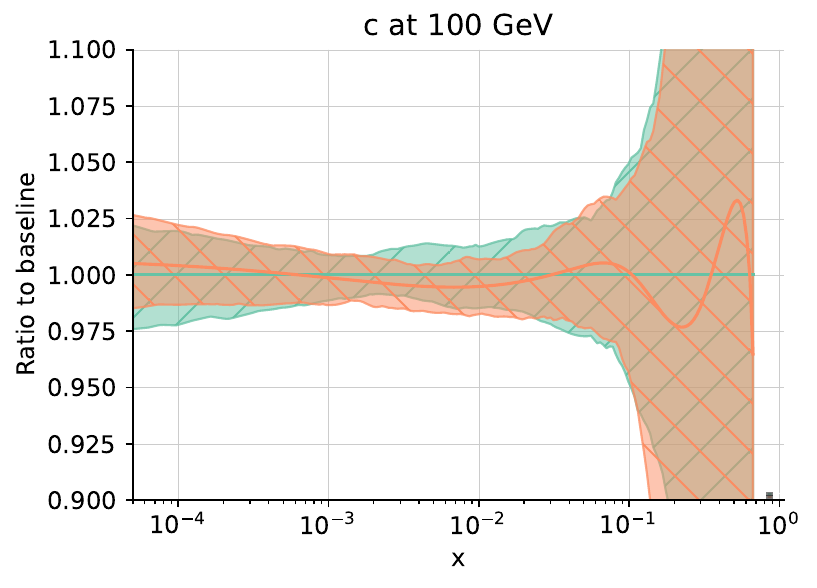}
  \includegraphics[width=0.48\textwidth]{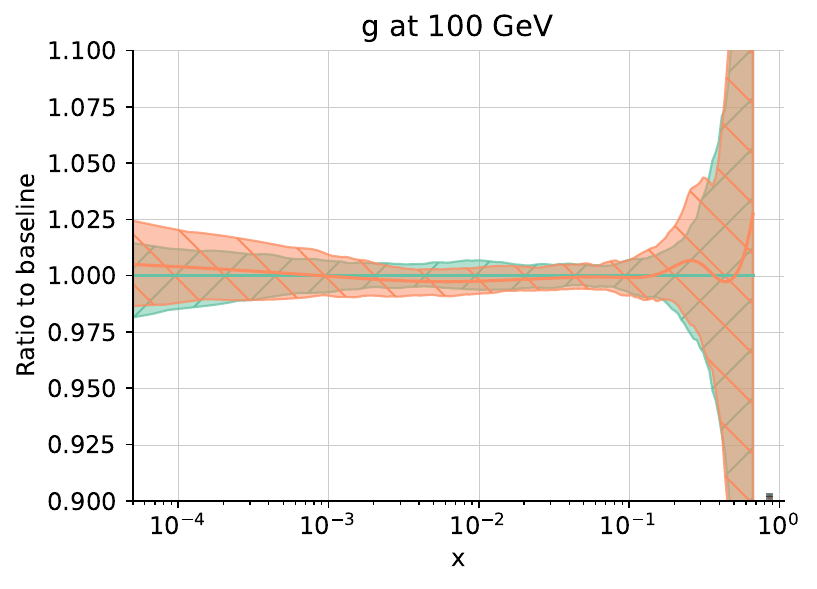}\\
  \caption{Comparison of the strange quark and anti-quark, charm quark and
    gluon PDFs obtained from the baseline and RES fits, see
    Table~\ref{tab:fits}. The format is as in
    Fig.~\ref{fig:PDFs_8TeV_baseline}.}
  \label{fig:PDFs_8TeV_baseline+resum}
\end{figure}

\begin{figure}[!t]
  \centering
  \includegraphics[width=0.48\textwidth]{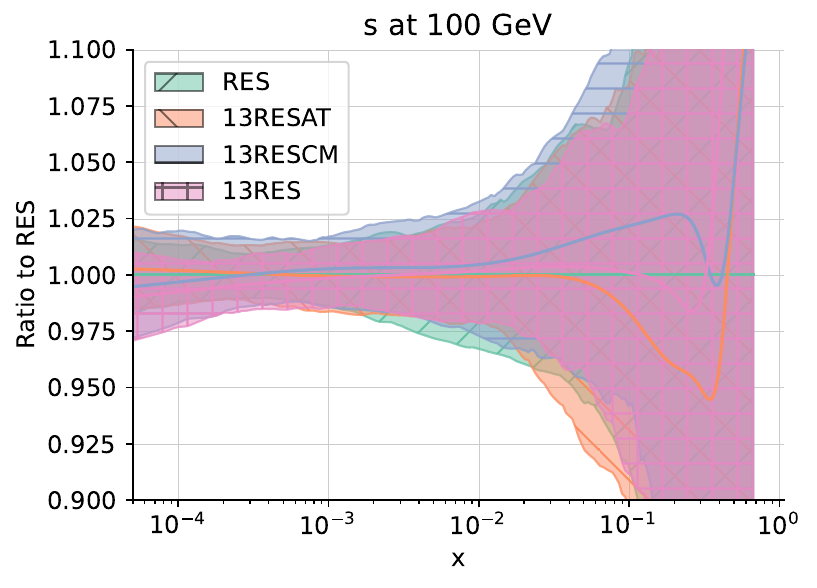}
  \includegraphics[width=0.48\textwidth]{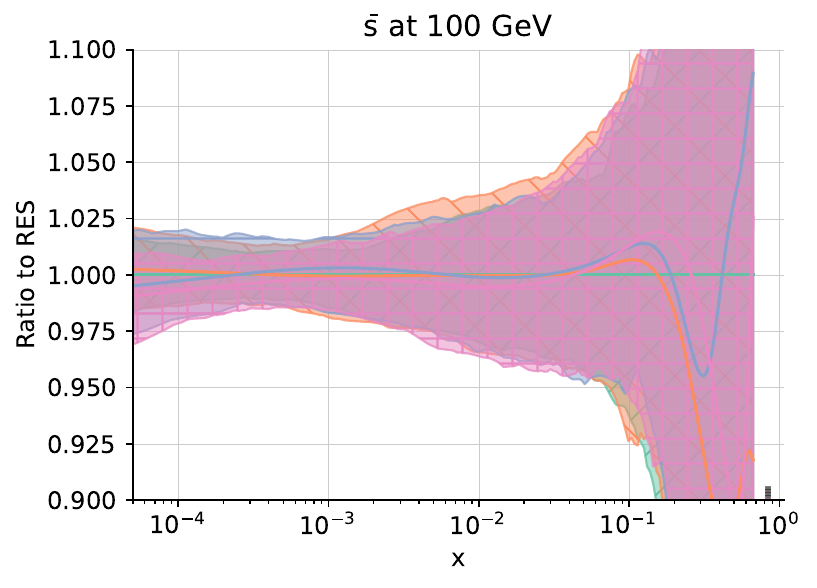}\\
  \includegraphics[width=0.48\textwidth]{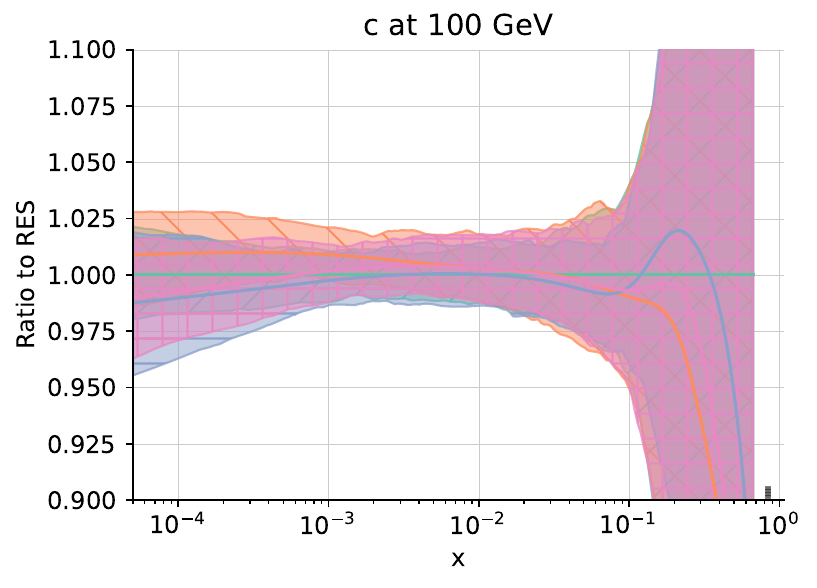}
  \includegraphics[width=0.48\textwidth]{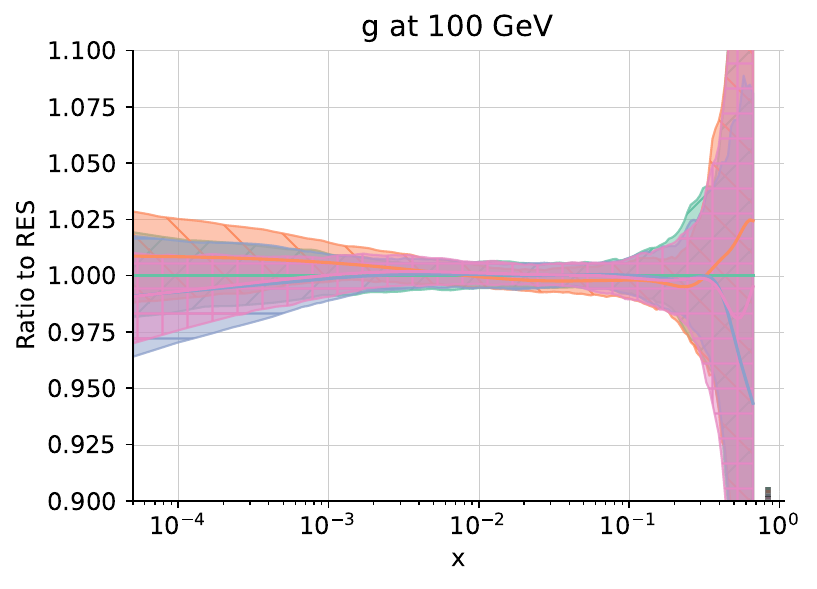}\\
  \caption{Comparison of the strange quark and anti-quark, charm quark and
    gluon PDFs obtained from the RES, 13RES, 13RESAT, and 13RESCM fits, see
    Table~\ref{tab:fits}. The format is as in
    Fig.~\ref{fig:PDFs_8TeV_baseline}.}
  \label{fig:PDFs_8TeV_baseline+resum+13TeV}
\end{figure}

We start by studying the case in which the cut is fixed and the $Z$-boson
transverse-momentum measurements included in the fits are varied. To this
purpose, we compare the baseline, RES, 13ATCM, 13RES, 13RESAT, and 13RESCM fits,
see Table~\ref{tab:fits}. In Table~\ref{tab:chi2_resummation_fixed_cut}, we
display the values of the $\csq$ per data point, $N_{\rm dat}$, for each of them.
We compare the strange quark and anti-quark, charm quark, and gluon PDFs
between the baseline and RES fits in Fig.~\ref{fig:PDFs_8TeV_baseline+resum},
and among the RES, 13RES, 13RESAT, and 13RESCM fits in
Fig.~\ref{fig:PDFs_8TeV_baseline+resum+13TeV}. The format of
Table~\ref{tab:chi2_resummation_fixed_cut}, and of
Figs.~\ref{fig:PDFs_8TeV_baseline+resum}
and~\ref{fig:PDFs_8TeV_baseline+resum+13TeV} is the same as
Table~\ref{tab:chi2_8TeV_baseline} and Fig.~\ref{fig:PDFs_8TeV_baseline},
respectively. Inspection of Table~\ref{tab:chi2_resummation_fixed_cut}
and Figs.~\ref{fig:PDFs_8TeV_baseline+resum},
\ref{fig:PDFs_8TeV_baseline+resum+13TeV}
leads us to the following conclusions.

%-------------------------------------------------------------------------------
\begin{table}[!p]
  \centering
  \footnotesize
  \renewcommand{\arraystretch}{1.2}
  \begin{tabularx}{\textwidth}{Xrcccccc}
    Data set
    & $N_{\rm dat}$
    & \rotatebox{90}{baseline}
    & \rotatebox{90}{RES}
    & \rotatebox{90}{13ATCM}
    & \rotatebox{90}{13RES}
    & \rotatebox{90}{13RESAT}
    & \rotatebox{90}{13RESCM}
    \\
    \toprule
    DIS NC
    & 2100 & 1.19 & 1.20 & 1.20 & 1.20 & 1.20 & 1.20 \\
    DIS CC
    &  989 & 0.89 & 0.89 & 0.89 & 0.89 & 0.89 & 0.89 \\
    DY NC
    &  671 & 1.23 & 1.19 & 1.44 & 1.25 & 1.28 & 1.17 \\
    \quad ATLAS $\ptll$ 8~TeV ($\ptll,\yll$)
    &   44 & 1.24 & 1.23 & 1.17 & 1.21 & 1.24 & 1.22\\
    \quad ATLAS $\ptll$ 8~TeV ($\ptll,\mll$)
    &   48 & 1.97 & 1.62 & 1.87 & 1.62 & 1.64 & 1.63\\
    \quad CMS $\ptll$ 8~TeV ($\ptll,\yll$)
    &   28 & 1.79 & 1.69 & 1.85 & 1.69 & 1.69 & 1.70\\
    \quad ATLAS $\ptll$ 13~TeV
    &   19 & ---  & ---  & 7.83 & 3.94 & 3.94 & ---\\
    \quad CMS $\ptll$ 13~TeV
    &   16 & ---  & ---  & 0.89 & 0.74 & ---  & 0.75\\
    DY CC
    &  157 & 1.07 & 1.06 & 1.05 & 1.06 & 1.04 & 1.08\\
    top-quark pair
    &   64 & 1.00 & 1.03 & 0.91 & 1.02 & 1.04 & 1.01\\
    single-inclusive jets
    &  356 & 0.87 & 0.88 & 0.89 & 0.89 & 0.89 & 0.87\\
    dijets
    &  144 & 1.52 & 1.50 & 1.51 & 1.51 & 1.51 & 1.50\\
    photon
    &   53 & 0.82 & 0.82 & 0.76 & 0.81 & 0.82 & 0.82\\
    single top-quark
    &   17 & 0.17 & 0.17 & 0.15 & 0.17 & 0.17 & 0.17\\
    \midrule
    Total
    & 4551 & 1.12 & 1.12 & 1.16 & 1.13 & 1.14 & 1.12\\
    \bottomrule
  \end{tabularx}
  \caption{The values of the $\csq$ per number of data points $N_{\rm dat}$
    for the baseline, RES, 13ATCM, 13RES, 13RESAT, and 13RESCM fits,
    see Table~\ref{tab:fits}. The format is the same as that of
    Table~\ref{tab:chi2_8TeV_baseline}. Note that the total number of data
    points does not include the data points in the 13~TeV ATLAS and CMS
    $Z$-boson transverse-momentum data sets.}
  \label{tab:chi2_resummation_fixed_cut}
\end{table}
%-------------------------------------------------------------------------------

%-------------------------------------------------------------------------------
%-------------------------------------------------------------------------------

%-------------------------------------------------------------------------------
%-------------------------------------------------------------------------------

We first compare the baseline and RES fits. We observe that the impact of
including small-$\ptll$ resummation corrections to the 8~TeV ATLAS and
CMS $Z$-boson transverse momentum measurements is very mild: their $\csq$
values improve modestly, decreasing from 1.24 to 1.23 and from 1.97 to 1.62
for the ATLAS measurements, and from 1.79 to 1.69 for the CMS measurements.
However, these improvements do not translate into an overall improvement of
the global fit quality, which remains essentially unchanged with respect
to the baseline fit. Parton distributions are consistently unmodified, except
for a very mild reduction of uncertainties. These results are coherent with the
shape and size of resummation $K$-factors, see Fig.~\ref{fig:res_K-factor}.

We then compare the RES, 13RES, 13RESAT and 13RESCM fits. In this case, we note
instead the significant impact of small-$\ptll$ resummation corrections.
At the level of the fit quality, we observe a substantial improvement of the
$\csq$ of the 13~TeV $Z$-boson transverse-momentum measurements, which
decreases from 7.83 to 3.94 for ATLAS, and from 0.89 to 0.74 for CMS. The
improvement in the description of the corresponding 8~TeV data, observed in the
RES fit, is maintained. These results are stable regardless of whether only the
ATLAS, only the CMS, or both ATLAS and CMS 13~TeV measurements are added to the
baseline data set. This leads to an overall $\csq$ of 1.13 in the 13RESCM fit,
which aligns pretty well with the $\csq$ of the starting baseline fit.

The relevance of small-$\ptll$ resummation corrections is even more
apparent at the level of PDFs. The strange quark and anti-quark, charm quark
and gluon PDFs obtained from the RES, 13RES, 13RESAT, and 13RESCM fit are all
compatible within uncertainties, see
Fig.~\ref{fig:PDFs_8TeV_baseline+resum+13TeV}. This contrasts with
Fig.~\ref{fig:PDFs_8TeV_baseline+13TeV_analyses}, where we observed
discrepancies among PDFs depending on the included data sets, specifically
due to the 13~TeV ATLAS $Z$-boson transverse-momentum measurement. Most notably,
the PDFs obtained in the 13RES fit are essentially identical to those obtained
in the RES fit, which is in turn very similar to the starting baseline fit,
see Fig.~\ref{fig:PDFs_8TeV_baseline+resum}.

All these observations lead us to two conclusions that we deem very relevant
for LHC phenomenology. First, small-$\ptll$ resummation corrections are
fundamental to ensure the compatibility of the very precise 13~TeV ATLAS
$Z$-boson transverse-momentum measurements with the baseline data set, even
when a relatively conservative $\ptll\geq 30$ GeV cut is applied to the data.
Second, provided that they are accompanied by resummation corrections, the
new $Z$-boson transverse-momentum measurements are unlikely to significantly
affect predictions for precision LHC processes, such as the gluon-fusion
Higgs cross section.
As noted, once small-$\ptll$ resummation corrections are included, only small
differences are observed among the baseline, RES, and 13RES PDF sets.

\subsubsection{Fixed data set}
\label{subsubsec:fixed_data_set}

\begin{figure}[!p]
  \centering
  \includegraphics[width=0.48\textwidth]{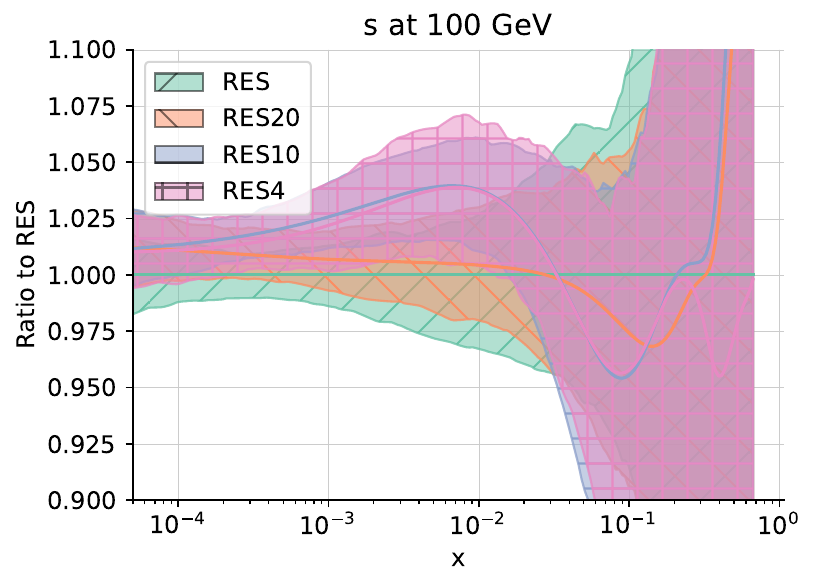}
  \includegraphics[width=0.48\textwidth]{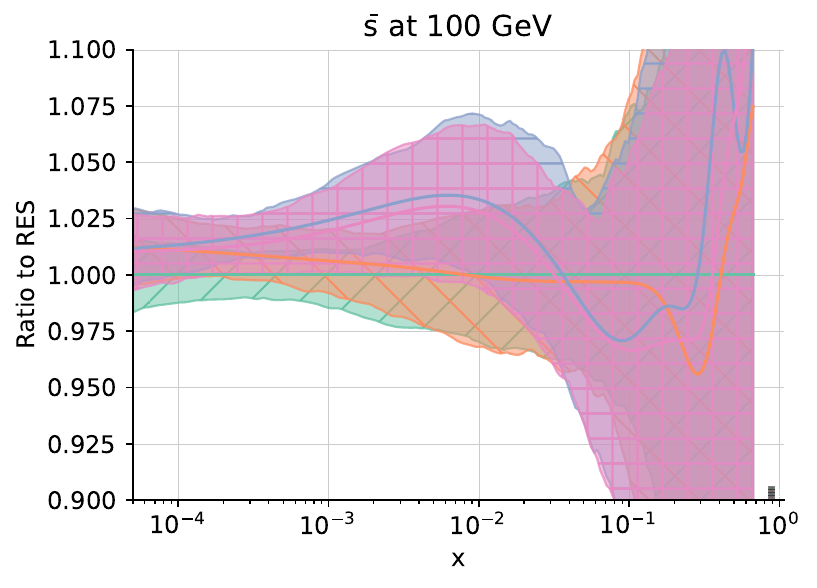}\\
  \includegraphics[width=0.48\textwidth]{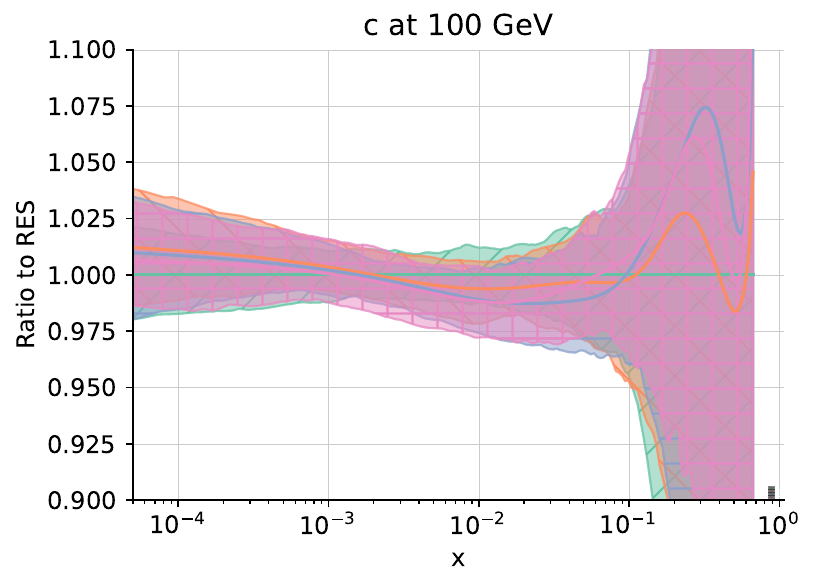}
  \includegraphics[width=0.48\textwidth]{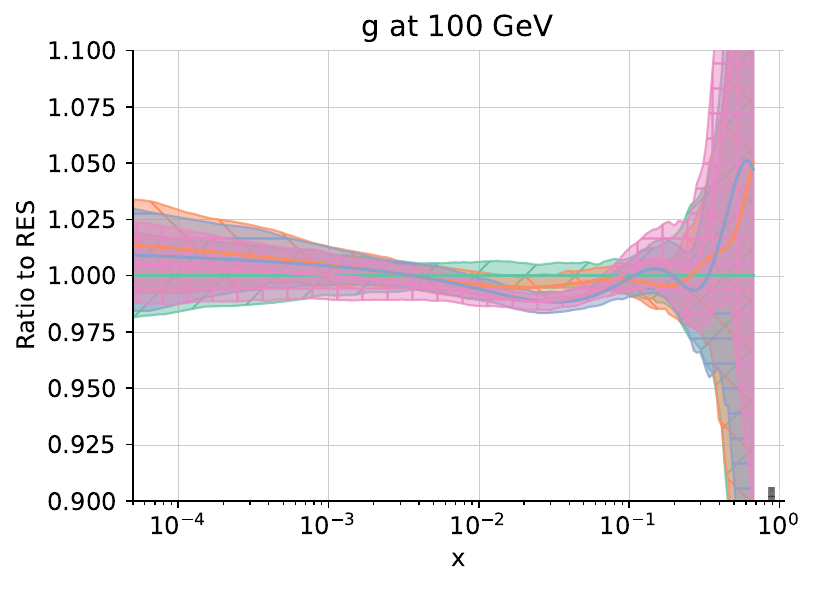}\\
  \caption{Comparison of the strange quark and anti-quark, charm quark, and
    gluon PDFs obtained from the RES, RES20, RES10, and RES4 fits, see
    Table~\ref{tab:fits}. The format is as in
    Fig.~\ref{fig:PDFs_8TeV_baseline}.}
  \label{fig:PDFs_8TeV_baseline+resum_low_cuts}
\end{figure}

\begin{figure}[!t]
  \centering
  \includegraphics[width=0.48\textwidth]{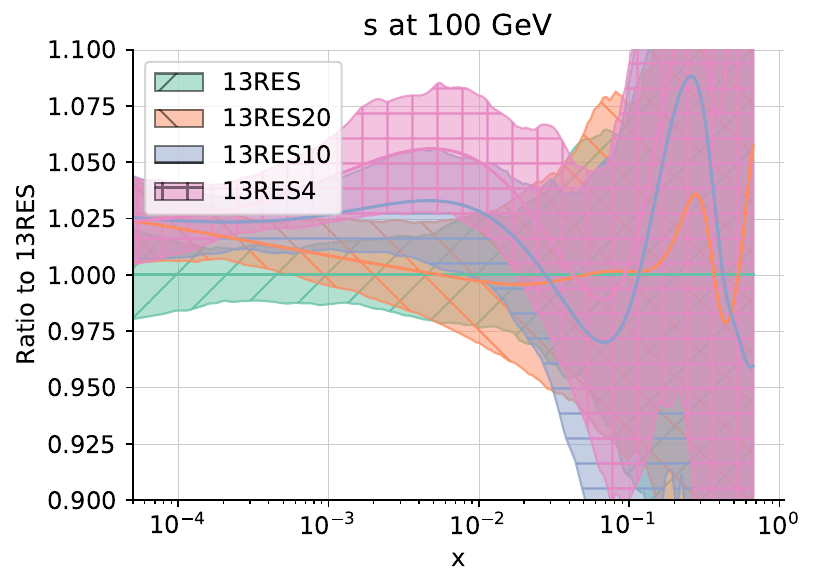}
  \includegraphics[width=0.48\textwidth]{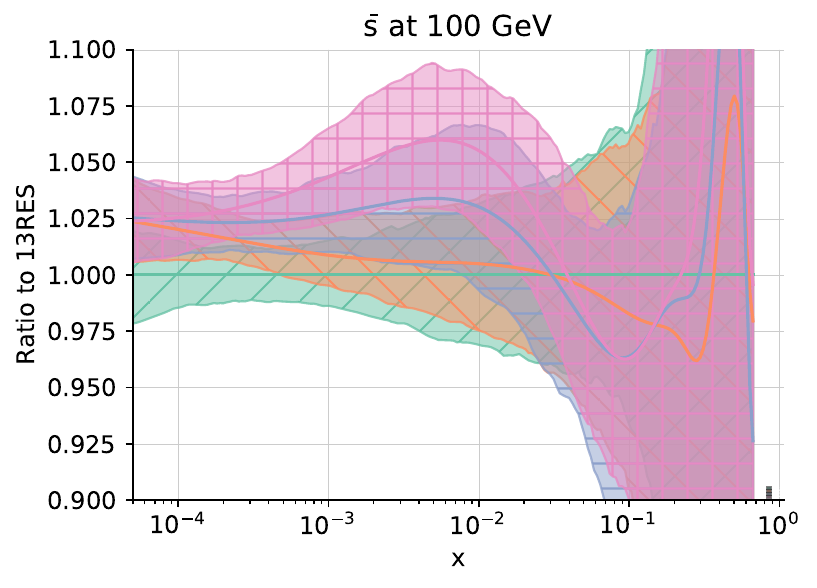}\\
  \includegraphics[width=0.48\textwidth]{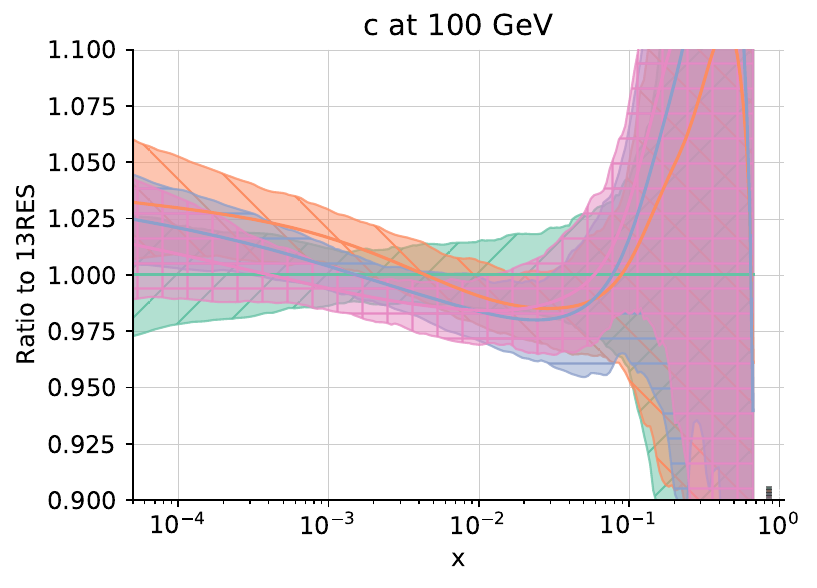}
  \includegraphics[width=0.48\textwidth]{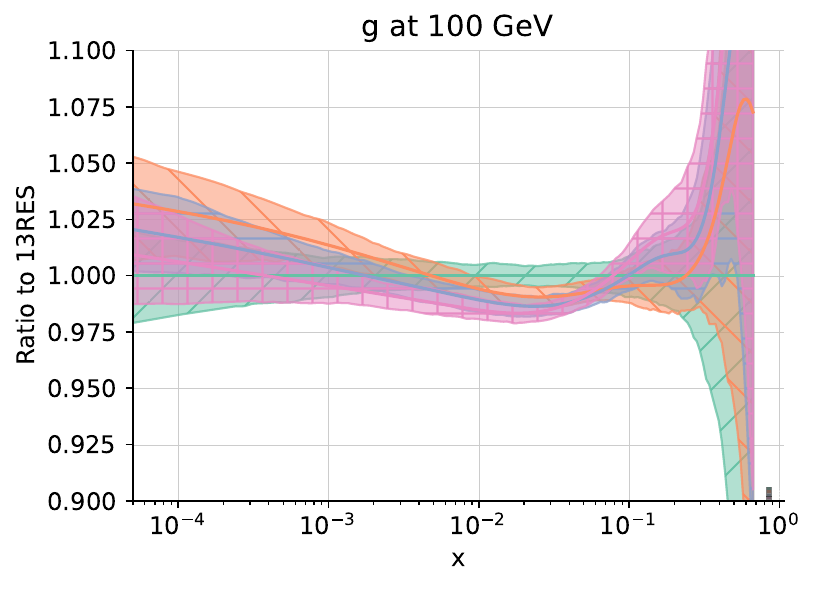}\\
  \caption{Comparison of the strange quark and anti-quark, charm quark, and
    gluon PDFs obtained from the 13RES, 13RES20, 13RES10, and 13RES4 fits, see
    Table~\ref{tab:fits}. The format is as in
    Fig.~\ref{fig:PDFs_8TeV_baseline}.}
  \label{fig:PDFs_8TeV_baseline+resum+13TeV_low_cuts}
\end{figure}

We then move to studying the case in which the cut $\ptmll$ on the $Z$-boson
transverse momentum measurements is lowered, while the data set is kept fixed.
As detailed in Table~\ref{tab:fits}, on top of the nominal value
$\ptmll=30$~GeV, we consider the additional values of 20, 10 and 4~GeV.
We consider two cases: one in which only the 8~TeV ATLAS and CMS
$Z$-boson transverse-momentum measurements are included in the fit, and one in
which both the 8 and 13~TeV ATLAS and CMS measurements are included.
In Table~\ref{tab:chi2_resummation_fixed_data_set}, we display the values of the
$\csq$ per data point, $N_{\rm dat}$, for each of these cases. We compare the
strange quark and anti-quark, charm quark, and gluon PDFs between the RES, RES20, RES10,
and RES4 fits in Fig.~\ref{fig:PDFs_8TeV_baseline+resum_low_cuts}, and between
the 13RES, 13RES20, 13RES10, and 13RES4 fits in
Fig.~\ref{fig:PDFs_8TeV_baseline+resum+13TeV_low_cuts}.
Inspection of Table~\ref{tab:chi2_resummation_fixed_data_set} and
Figs.~\ref{fig:PDFs_8TeV_baseline+resum_low_cuts},~\ref{fig:PDFs_8TeV_baseline+resum+13TeV_low_cuts}
leads us to the following observations.

%-------------------------------------------------------------------------------
\begin{table}[!p]
  \centering
  \footnotesize
  \renewcommand{\arraystretch}{1.2}
  \begin{tabularx}{\textwidth}{Xrcccccccc}
    Data set
    & $N_{\rm dat}$
    & \rotatebox{90}{RES}
    & \rotatebox{90}{RES20}
    & \rotatebox{90}{RES10}
    & \rotatebox{90}{RES4}
    & \rotatebox{90}{13RES}
    & \rotatebox{90}{13RES20}
    & \rotatebox{90}{13RES10}
    & \rotatebox{90}{13RES4}\\
    \toprule
    DIS NC
    & 2100 & 1.20 & 1.20 & 1.20 & 1.21
           & 1.20 & 1.21 & 1.21 & 1.21 \\
    DIS CC
    &  989 & 0.89 & 0.89 & 0.89 & 0.90
           & 0.89 & 0.89 & 0.90 & 0.90 \\
    DY NC
    &  671 & 1.19 & 1.41 & 1.74 & 1.90
           & 1.25 & 1.62 & 2.29 & 2.64 \\
    \, ATLAS~$\ptll$~8~TeV~($\ptll,\yll$)
    &   44 & 1.23 & 1.24 & 1.24 & 1.22
           & 1.21 & 1.27 & 1.23 & 1.21 \\
    \, ATLAS~$\ptll$~8~TeV~($\ptll,\mll$)
    &   48 & 1.62 & 3.71 & 3.78 & 4.40
           & 1.62 & 3.65 & 3.56 & 4.31 \\
    \, CMS $\ptll$ 8~TeV ($\ptll,\yll$)
    &   28 & 1.69 & 1.65 & 4.47 & 4.42
           & 1.69 & 1.64 & 4.64 & 4.53 \\
    \, ATLAS $\ptll$ 13~TeV
    &   19 & ---  & ---  & ---  & ---
           & 3.94 & 8.42 & 13.5 & 17.6 \\
    \, CMS $\ptll$ 13~TeV
    &   16 & ---  & ---  & ---  & ---
           & 0.74 & 0.80 & 1.03 & 1.63 \\
    DY CC
    &  157 & 1.06 & 1.09 & 1.09 & 1.06
           & 1.06 & 1.09 & 1.04 & 1.02 \\
    top-quark pair
    &   64 & 1.03 & 1.04 & 1.04 & 1.21
           & 1.02 & 1.13 & 1.26 & 1.29 \\
    single-inclusive jets
    &  356 & 0.88 & 0.88 & 0.88 & 0.86
           & 0.98 & 0.98 & 0.95 & 0.96 \\
    dijets
    &  144 & 1.50 & 1.52 & 1.52 & 1.51
           & 1.51 & 1.51 & 1.48 & 1.46 \\
    photon
    &   53 & 0.82 & 0.83 & 0.83 & 0.88
           & 0.81 & 0.83 & 0.86 & 0.85 \\
    single top-quark
    &   17 & 0.17 & 0.18 & 0.18 & 0.18
           & 0.17 & 0.18 & 0.18 & 0.18 \\
    \midrule
    Total
    & 4551 & 1.12 & 1.15 & 1.21 & 1.24
           & 1.13 & 1.20 & 1.31 & 1.39 \\
    \bottomrule
  \end{tabularx}
  \caption{The values of the $\csq$ per number of data points $N_{\rm dat}$
    for the RES, RES20, RES10, RES4, 13RES, 13RES20, 13RES10 and 13RES4
    fits, see Table~\ref{tab:fits}.  The format is as in
    Table~\ref{tab:chi2_8TeV_baseline}. Note that the total number of data
    points does not include the data points in the 13~TeV ATLAS and CMS
    $Z$-boson transverse momentum data sets.}
  \label{tab:chi2_resummation_fixed_data_set}
\end{table}
%-------------------------------------------------------------------------------

%-------------------------------------------------------------------------------
%-------------------------------------------------------------------------------

%-------------------------------------------------------------------------------
%-------------------------------------------------------------------------------

Relaxing the $\ptll$ cut generally results in a deterioration of the
$\csq$ of all the $Z$-boson transverse-momentum measurements, irrespective
of their centre-of-mass energy (8 or 13~TeV), and of the experiment (ATLAS or
CMS). The deterioration is continuous, in that it gets larger
as the cut is lowered. Interestingly, we see a concurrent deterioration of the
$\csq$ of top-quark pair and single-inclusive jet measurements, particularly
for the lowest $\ptll$ cut when including the 13~TeV $Z$-boson
transverse-momentum measurements. The $\csq$ of the other data sets is generally
stable upon variations of the $\ptll$ cut. At the level of PDFs, we
observe similar variations in the fits that include only the 8~TeV ATLAS and
CMS $Z$-boson transverse-momentum measurements or both the 8 and 13~TeV ones,
as the  $\ptll$ cut is lowered. The effect is an enhancement in the
central value of the strange quark and anti-quark PDFs around $x\sim 0.01$, and
a distortion of the shape of the charm quark and gluon PDFs: an enhancement for
$x\lesssim 0.001$ and a depletion around $x\sim 0.03$.
The effect is milder in fits including only the 8~TeV data, for which the
aforementioned shifts never exceed the nominal PDF uncertainty. In contrast,
it is more pronounced in fits including both the 8 and 13~TeV data, where the
shifts in the gluon PDF can reach about 2.5 sigma when expressed in units of
the PDF uncertainty. In light of these findings, we consider it prudent to
retain the more conservative $\ptll\geq 30$~GeV cut in PDF fits. More
generally, these results motivate three further considerations.

The first consideration concerns the appropriateness of our treatment of
small-$\ptll$ corrections. As described in Sect.~\ref{subsubsec:resumm},
these are incorporated in our fitting framework by means of resummation
$K$-factors. This approach is an approximation, in that $K$-factors rescale
the cross section by a bin-dependent multiplicative correction that does not
reweigh individual partonic channels. Testing the faithfulness of this
approximation would require to include resummation corrections in an exact
form, for instance by interfacing \rad to {\sc PineAPPL}. In this way, the
weights stored in the interpolation grid would account not only for the
fixed-order, but also for the resummed computation. This development is 
left for future work. We note however that, in the case of NNLO QCD
corrections, the $K$-factor approximation for DY data has been demonstrated to provide
an accurate description, and to deteriorate predictions only
marginally~\cite{Cruz-Martinez:2025ffa}.

The second consideration concerns the possible impact of (approximate)
N$^3$LO corrections in the PDF fit. Such corrections affect the evolution
equations and DIS coefficient functions, including massive terms, properly
matched in a general-mass variable flavour number scheme.
They do not encompass coefficient functions for LHC processes, whose
N$^3$LO corrections are mostly unknown. Nevertheless, approximate N$^3$LO PDF
fits exist~\cite{McGowan:2022nag,NNPDF:2024nan,Cridge:2024icl}, which reveal
a suppression of the gluon PDF by 1-2\% around $x\sim 0.03$. Such an effect is
very similar to that observed in
Figs.~\ref{fig:PDFs_8TeV_baseline+resum_low_cuts}
and~\ref{fig:PDFs_8TeV_baseline+resum+13TeV_low_cuts}.
One may therefore wonder whether our conclusions would change if the
analysis were repeated at approximate N$^3$LO. We note that this exercise
could be performed only approximately, since N$^3$LO fixed-order predictions
for the $\ptll$ spectrum are not yet available.
Nevertheless, we have recomputed the $\csq$ with fixed PDFs, by using now the
NNPDF4.0 N$^3$LO sets without~\cite{NNPDF:2024nan} and
with~\cite{Barontini:2024dyb} QED corrections, which may further suppress the
gluon PDF. The $\csq$ values do not change significantly in comparison to those
reported in Table~\ref{tab:chi2_resummation_fixed_data_set}. This fact suggests
that, even if we repeated our analysis at approximate N$^3$LO, we would
not find significantly different results.

%-------------------------------------------------------------------------------
\begin{table}[!t]
  \centering
  \footnotesize
  \renewcommand{\arraystretch}{1.2}
  \begin{tabularx}{\textwidth}{Xrcccccccc}
    Data set
    & $N_{\rm dat}$
    & \rotatebox{90}{13RES}
    & \rotatebox{90}{13RES20}
    & \rotatebox{90}{13RES10}
    & \rotatebox{90}{13RES4}
    & \rotatebox{90}{13RES}
    & \rotatebox{90}{13RES20}
    & \rotatebox{90}{13RES10}
    & \rotatebox{90}{13RES4}\\
    \toprule
    &
    & \multicolumn{4}{c}{diagonal prescription}
    & \multicolumn{4}{c}{envelope prescription}\\
    \midrule
    ATLAS~$\ptll$~8~TeV~($\ptll,\yll$)
    &   44 & 0.62 & 0.64 & 0.63 & 0.61
           & 0.53 & 0.55 & 0.54 & 0.53 \\
    ATLAS~$\ptll$~8~TeV~($\ptll,\mll$)
    &   48 & 0.39 & 0.44 & 0.39 & 0.35
           & 0.28 & 0.32 & 0.29 & 0.26 \\
    CMS $\ptll$ 8~TeV ($\ptll,\yll$)
    &   28 & 0.30 & 0.37 & 0.60 & 0.71
           & 0.18 & 0.24 & 0.39 & 0.47 \\
    ATLAS $\ptll$ 13~TeV
    &   19 & 0.39 & 0.42 & 0.49 & 0.47
           & 0.29 & 0.32 & 0.35 & 0.33 \\
    CMS $\ptll$ 13~TeV
    &   16 & 0.61 & 0.73 & 0.71 & 0.75
           & 0.52 & 0.62 & 0.63 & 0.70 \\
    \bottomrule
  \end{tabularx}
  \caption{The values of the $\csq$ per number of data points $N_{\rm dat}$
    for the 13RES, 13RES20, 13RES10 and 13RES4 PDFs, see Table~\ref{tab:fits},
    recomputed with the diagonal and envelope prescriptions for the computation
    of the theory covariance matrix, see text for details. We display results
    only for the $Z$-boson transverse-momentum measurements.}
  \label{tab:theory_cov_off}
\end{table}
%-------------------------------------------------------------------------------

%-------------------------------------------------------------------------------
\begin{figure}[!t]
  \includegraphics[width=0.48\textwidth]{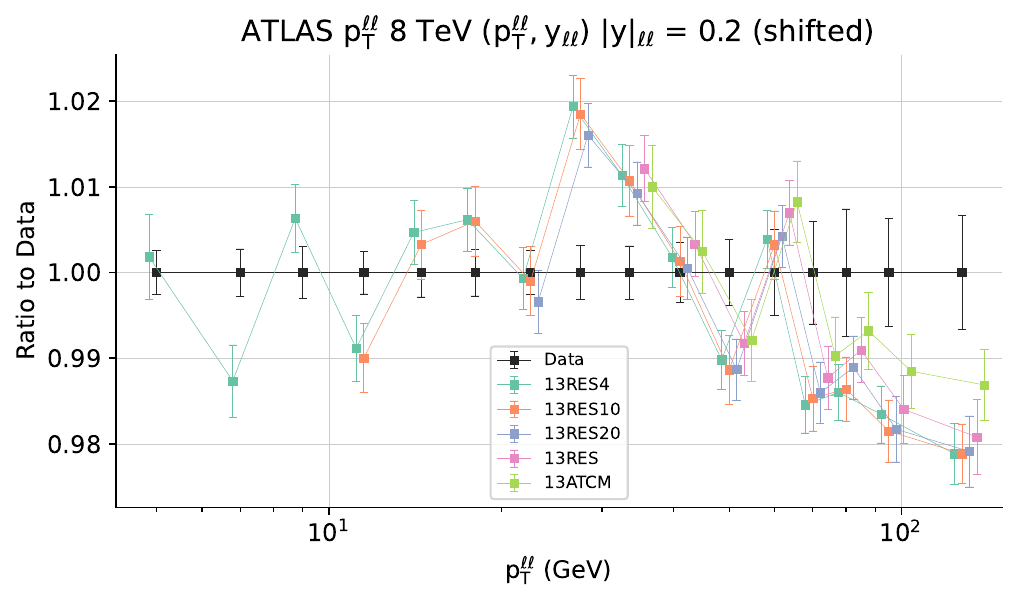}
  \includegraphics[width=0.48\textwidth]{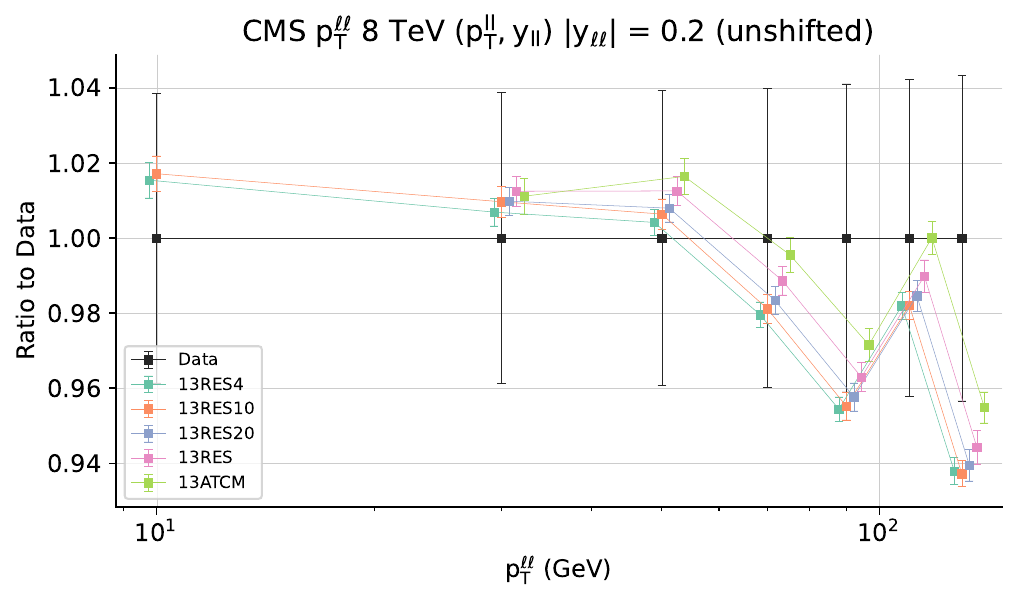}\\
  \includegraphics[width=0.48\textwidth]{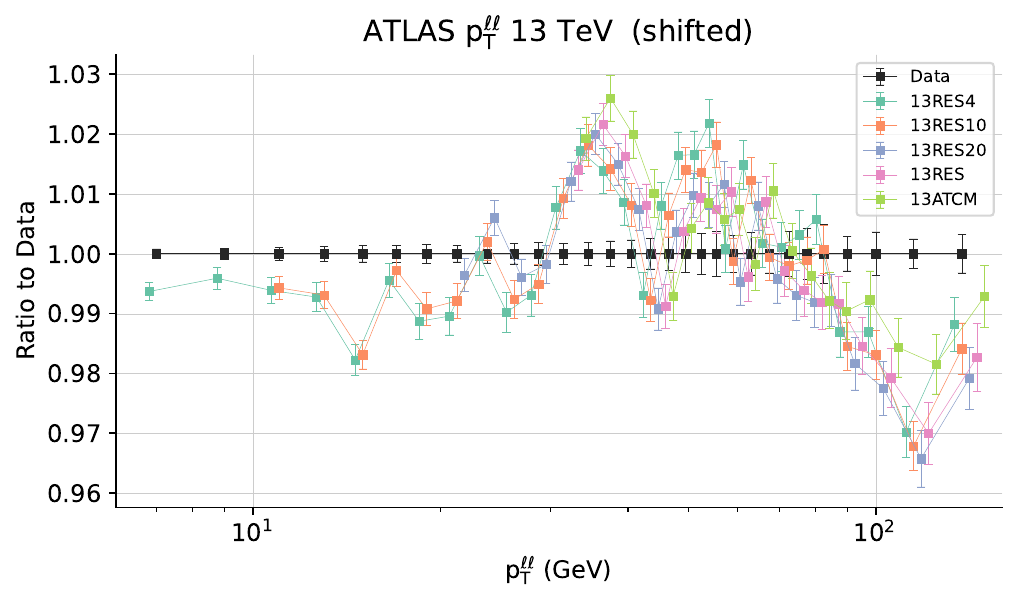}
  \includegraphics[width=0.48\textwidth]{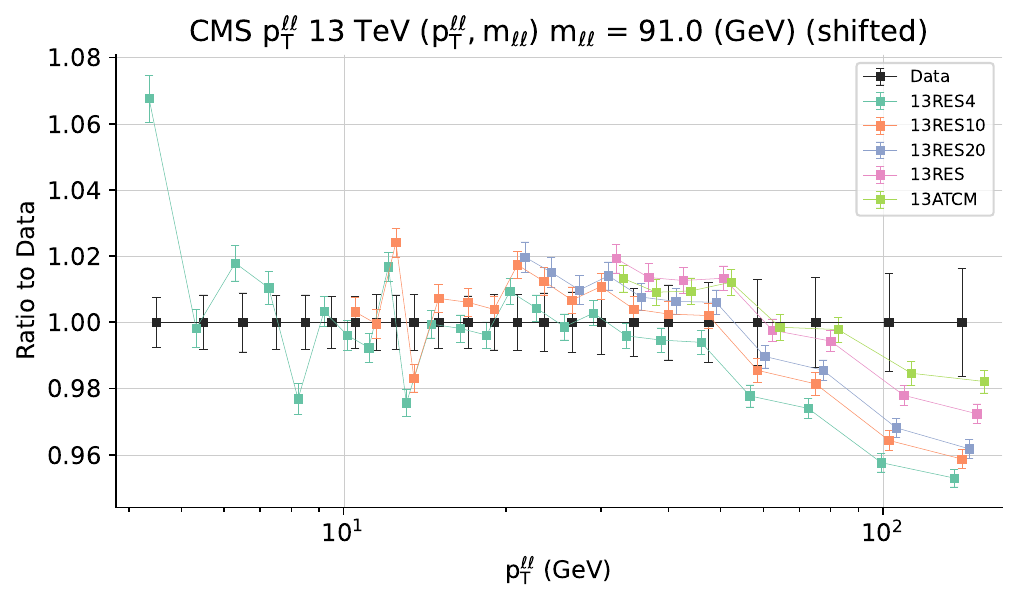}\\  
  \caption{Data-theory comparison plots for a selection of $Z$-boson
    transverse-momentum measurements. From top to bottom, left to right, we
    show: the $0.0\leq \yll\leq 0.4$ lepton-pair rapidity bin of the ATLAS 8~TeV
    measurement; the $0.0 \leq \yll\leq 0.4$ lepton-pair rapidity bin of
    the CMS 8~TeV measurement; the ATLAS 13~TeV measurement; and the $Z$-peak
    invariant-mass bin of the CMS 13~TeV measurement. In all plots we display
    predictions corresponding to the 13ATCM, 13RES, 13RES20, 13RES10, and 13RES4
    fits. To improve the readability of the figure, theoretical predictions are
    shifted, if possible, by an amount that depends on the correlation of
    experimental uncertainties, see text for details. The uncertainty on the
    data corresponds to the sum in quadrature of all the uncorrelated
    experimental uncertainties; the uncertainty on theoretical predictions is
    the PDF uncertainty only.}
  \label{fig:data_theory}
\end{figure}
%-------------------------------------------------------------------------------

%-------------------------------------------------------------------------------
\begin{figure}[!t]
  \includegraphics[width=0.48\textwidth]{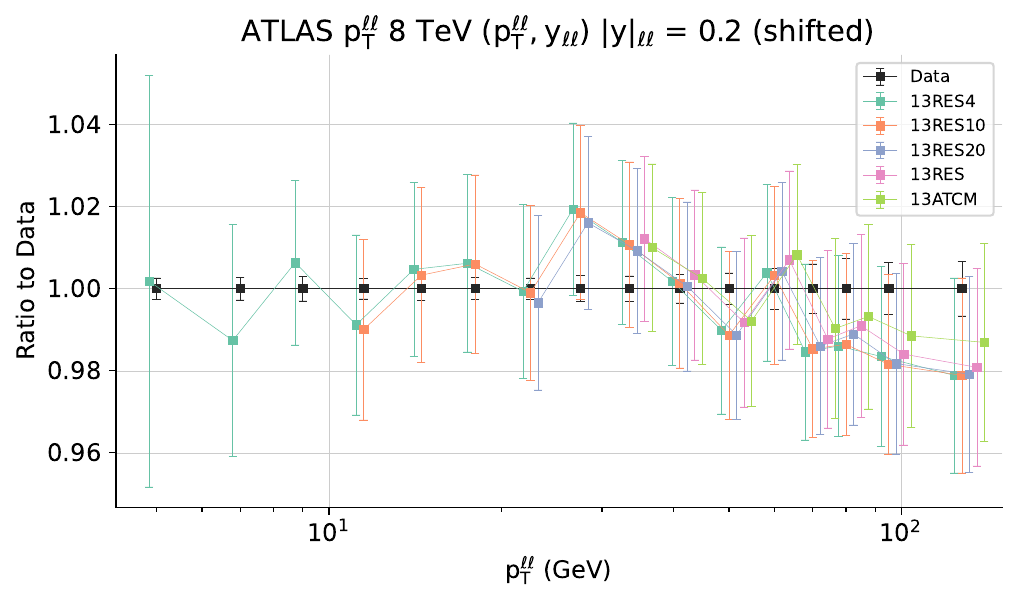}
  \includegraphics[width=0.48\textwidth]{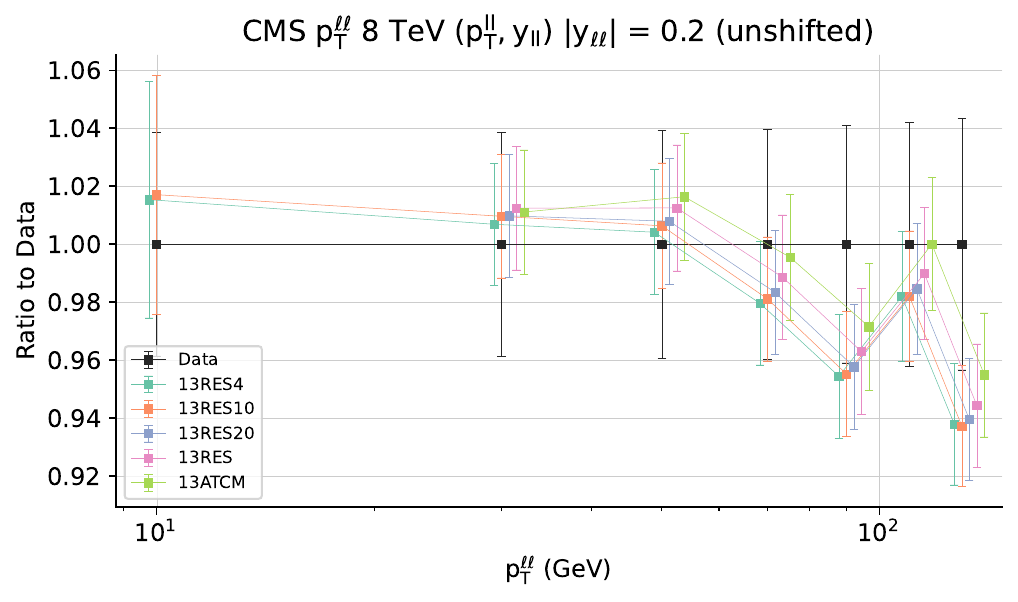}\\
  \includegraphics[width=0.48\textwidth]{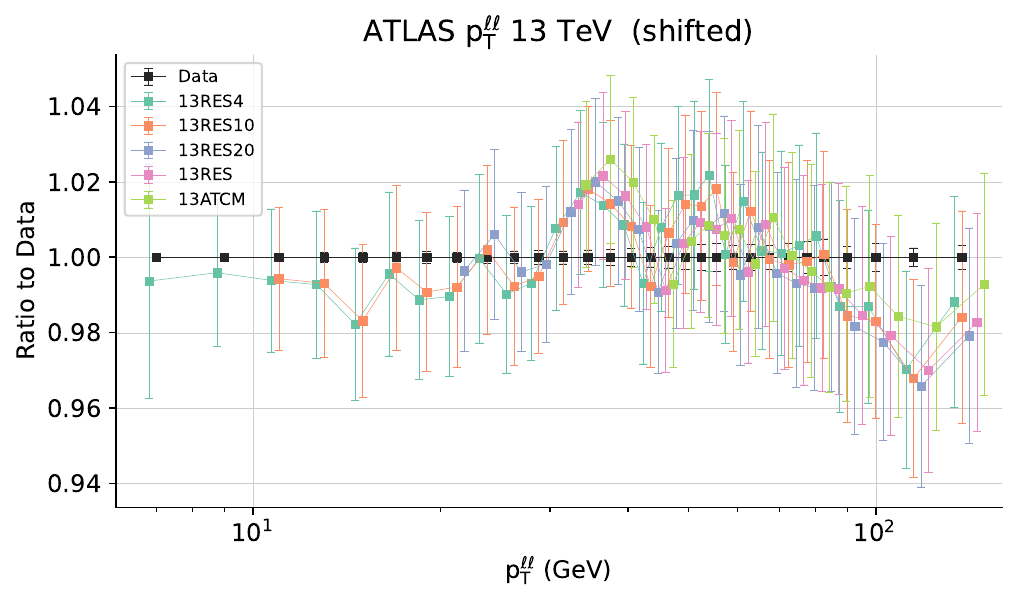}
  \includegraphics[width=0.48\textwidth]{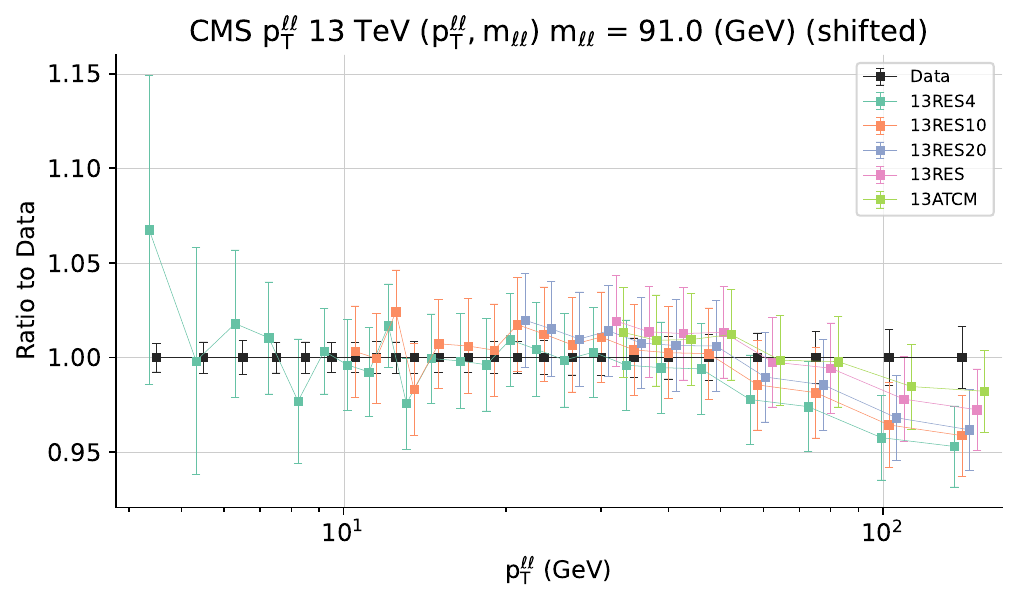}\\
  \caption{Same as Fig.~\ref{fig:data_theory}, with uncertainties on theoretical
    predictions equal to the sum in quadrature of the PDF uncertainty and of the
    diagonal elements of the theory covariance matrix.}
  \label{fig:data_theory_sv}
\end{figure}
%-------------------------------------------------------------------------------

The third and final consideration is that there may not exist a unique scale
variation preferred by the fit across the entire $\ptll$ spectrum. Rather, the
preferred scale variations may differ between the low- and high-$\ptll$
regions of the $Z$-boson transverse-momentum measurements.
This fact may be responsible for the large values of the $\csq$ reported in
Table~\ref{tab:chi2_resummation_fixed_data_set}, if different scale variations
are overly correlated in the theory covariance matrix.
This conjecture stems from the fact that resummed predictions are known to
provide a very good description of experimental DY data down to very small
$\ptll$ values, see for instance~\cite{ATLAS:2023lsr}. The same applies when
data are described in the framework of transverse-momentum-dependent
PDFs in the region $2\lesssim \ptll\lesssim 10$~GeV~\cite{Bacchetta:2022awv,
  Bacchetta:2024qre,Bacchetta:2025ara,Moos:2023yfa,Moos:2025sal}, although in
fits that contain a more limited number of data sets in comparison to the fits
of collinear PDFs presented in this work. This is in apparent contrast with the
large $\csq$ values reported in Table~\ref{tab:chi2_resummation_fixed_data_set}
for low-$\ptll$ cuts. We then wonder whether our treatment of missing
higher-order uncertainties, which consists in correlating renormalisation- and
factorisation-scale variations across data sets, see
Sect.~\ref{subsubsec:th_cov}, is not overconstraining the actual theory
uncertainty.

To check this aspect, we recompute the $\csq$ for the $Z$-boson
transverse-momentum measurements with the settings and PDFs of the 13RES,
13RES20, 13RES10, and 13RES4 fits, but employing a different prescription for
the treatment of theory uncertainties.
Specifically, we consider a {\it diagonal} scenario, in which all off-diagonal
elements of the theory covariance matrix are set to zero. We also define an
{\it envelope} scenario, in which the theory covariance matrix is taken to be
diagonal, with diagonal elements given by the differences between the
outermost scale-varied predictions.
In both cases, we consider 7-point scale variations, as in the default setup
where we retain correlations. The diagonal and envelope prescriptions are likely
to overestimate missing higher-order uncertainties, hence we use them only
as a diagnostic tool to check whether and how much the $\csq$ can reduce.
We report our results in Table~\ref{tab:theory_cov_off}, which we comment
upon below.

To complement this analysis, in Figs.~\ref{fig:data_theory}
and~\ref{fig:data_theory_sv} we collect data-theory comparison plots for a
selection of $Z$-boson transverse-momentum measurements. Specifically, we
show: the $0.0\leq \yll\leq 0.4$ lepton-pair rapidity bin of the
ATLAS 8~TeV measurement; the $0.0\leq \yll\leq 0.4$ lepton-pair rapidity
bin of the CMS 8~TeV measurement; the ATLAS 13~TeV measurement; and the $Z$-peak
invariant-mass bin of the CMS 13~TeV measurement. In all plots we display
predictions corresponding to the 13ATCM, 13RES, 13RES20, 13RES10, and 13RES4
fits. To improve the readability of the figure, theoretical predictions
are shifted by an amount that depends on the correlation of experimental
uncertainties, see Appendix~B in~\cite{Pumplin:2002vw}
for details.\footnote{This is not possible for the 8~TeV CMS measurement,
for which only the full experimental covariance matrix is provided, with
no separation between its uncorrelated and correlated components.} In both
Figs.~\ref{fig:data_theory} and~\ref{fig:data_theory_sv}, the uncertainty on
the data corresponds to the sum in quadrature of all the uncorrelated
experimental uncertainties.
As for uncertainties on theory predictions, in Fig.~\ref{fig:data_theory} we
display only the PDF uncertainties, the reason being that the effect of
correlated theory uncertainties is already taken into account by the fit as a
shift of the central value of the predictions. Figure~\ref{fig:data_theory}
should then be assessed in conjunction with the $\csq$ values of
Table~\ref{tab:chi2_resummation_fixed_data_set}.
Conversely, in Fig.~\ref{fig:data_theory_sv} uncertainties on theory predictions
are the sum in quadrature of the PDF uncertainty and of the diagonal elements
of the theory covariance matrix. This corresponds to the diagonal prescription
defined above and can be compared to the $\csq$ values displayed in
Table~\ref{tab:theory_cov_off}.
In all plots, the results are normalised to the central value of the data.

Comparing Fig.~\ref{fig:data_theory} with Fig.~\ref{fig:data_theory_sv}, we
conclude that experimental data and theoretical predictions are in agreement
within the corresponding uncertainties only if one does not take into account
correlations of theory uncertainties, as done in Fig.~\ref{fig:data_theory_sv}.
Neglecting correlations is of course incorrect: the values of the $\csq$ in
Table~\ref{tab:theory_cov_off} are obviously anomalously small, and theory
uncertainties in Fig.~\ref{fig:data_theory_sv} are likely to be
overestimated.
The opposite strategy, namely that of fully correlating renormalisation- and
factorisation-scale uncertainties across data points, as done through the fit
in Fig.~\ref{fig:data_theory}, may however be too aggressive, as it neglects
the possibility that different perturbative
behaviours are preferred by the data in different regions of $\ptll$.
Be that as it may, we believe that these results indicate that the treatment
of theory uncertainties, and in particular of their correlations across
different regions of $\ptll$, deserves careful consideration and may warrant
reassessment in future work, potentially employing different theory-uncertainty
estimators~\cite{Tackmann:2024kci}.

%% file: sec-conclusions.tex
\section{Conclusions}
\label{sec:conclusions}

In this paper, we have investigated the impact of $Z$-boson transverse-momentum
($\ptll$) measurements in the DY process on the determination of proton PDFs
within the NNPDF global fitting methodology. Our analysis combines NNLO QCD
predictions for the relevant processes with small-$\ptll$ resummation
corrections for NC DY production; it supplements the experimental covariance
matrix with a theory covariance matrix that estimates missing higher-order
uncertainties from renormalisation- and factorisation-scale variations.
The focus has been on the ATLAS and CMS measurements of the $\ptll$ spectrum at
8 and 13~TeV, with special emphasis on the possibility of extending the fitted
kinematic region to significantly lower values of $\ptll$ than those customarily
used in PDF fits. Our main findings can be summarised as follows.

First, we have reassessed the treatment of the 8~TeV $Z$-boson
transverse-momentum data already present in the NNPDF4.0 data set. The use of
exact NNLO interpolation grids, a uniform choice of central scales for DY data,
full-colour dijet predictions, and \nnlojet predictions for the $Z$-boson
transverse-momentum distributions leads to a more robust theoretical baseline.
In particular, the improved numerical stability of the \nnlojet calculation
allows us to remove the additional uncorrelated $1\%$ uncertainty previously
assigned to the 8~TeV $\ptll$ data, without inducing significant
distortions in the resulting PDFs.

Second, we have quantified the impact of the new 13~TeV $Z$-boson
transverse-momentum measurements. When these data are described with fixed-order
predictions only, the CMS measurement is well accommodated in the global fit,
while the ATLAS measurement is described rather poorly and induces visible
shifts in the PDFs, most notably in the strange and gluon distributions. This
points to a tension between the fixed-order description and the high precision
of the ATLAS 13~TeV data, rather than to a sizeable new PDF constraint that can
be easily absorbed by the fit.

Third, we have shown that small-$\ptll$ resummation corrections play a
central role in resolving this situation. With the standard cut
$\ptll\geq 30$~GeV, resummation has a mild but visible effect on the 8~TeV data,
improving their description while leaving PDFs essentially unchanged apart from
a modest reduction of uncertainties. The impact of resummation
is more pronounced for the 13~TeV data: the description of the ATLAS measurement
improves substantially, the CMS description is maintained or slightly improved,
and the PDFs obtained after including the new data remain compatible with the
baseline fit. This indicates that the apparent PDF shifts induced by the
13~TeV data at fixed order are largely mitigated once the appropriate
logarithmic corrections are included.

Fourth, we have investigated whether the inclusion of resummed predictions
permits lowering the minimum-$\ptll$ cut below its conventional value of
30~GeV. Our results show that extending the fit region to smaller
transverse momenta is in principle possible, but not uniformly successful for
all experiments and all bins under the present treatment of theory
uncertainties. The deterioration observed when the cut is lowered suggests that
the correlation model adopted for theoretical missing higher-order uncertainties
is a critical ingredient. While neglecting correlations among theory
uncertainties would artificially improve the description of the data, a fully
correlated treatment of scale variations across kinematic regions may be too
restrictive, especially if different regions of the $\ptll$ spectrum prefer
different perturbative behaviours.

These observations lead to a broader conclusion. Percent- and sub-percent-level
measurements of differential DY spectra cannot be fully exploited in PDF
determinations without theory predictions of commensurate accuracy, including
both all-order resummation in the appropriate regions of phase space, and a
careful treatment of missing higher-order uncertainties and their correlations.
The resummation-scale variation, which has not been included in the theory
covariance matrix because of its comparatively small impact in the present
study, should be revisited as part of a more comprehensive assessment of
theoretical uncertainties at low $\ptll$.

The results of this work therefore support a systematic inclusion of resummed
calculations in future global PDF analyses. Such an approach is necessary not
only to enlarge the kinematic reach of fitted DY data, but also to
avoid attributing deficiencies of fixed-order theory to spurious PDF effects.
Furthermore, the determination of $\as$ from global PDF analyses would also
benefit from the inclusion of data at small transverse momentum, due to the
high sensitivity of the peak of the $\ptll$ distribution to the value of the
strong coupling \cite{Camarda:2022qdg,ATLAS:2023lhg}.
As experimental uncertainties continue to decrease, the consistent combination
of high-precision measurements, resummed perturbative predictions, and reliable
theory covariance prescriptions will become increasingly important for
precision LHC phenomenology. The PDF sets discussed in this paper are available,
in the LHAPDF format~\cite{Buckley:2014ana}, from the authors upon request.

%% file: acknowledgements.tex
\section*{Acknowledgements}

We thank the members of the NNPDF collaboration for discussions and support.
We are grateful to Pier Francesco Monni and Emanuele Re for useful comments
on the manuscript. J.C.-M. acknowledges support from the Ramón y Cajal program
grant RYC2023-043794-I funded by MCIN/AEI/ 10.13039/501100011033 and by ESF+. 
The work of E.R.N. was supported by the Italian Ministry of University and
Research (MUR) through the “Rita Levi-Montalcini” Program.
The work of P.T. has been partially supported by the Italian MUR through
grant PRIN 2022BCXSW9, and by Compagnia di San Paolo through grant
TORP\_S1921\_EX-POST\_21\_01.

%% file: app-weighted_fit.tex
\section{The 13~TeV ATLAS $Z$-boson transverse-momentum measurement}
\label{app:weighted_fit}

In this Appendix, we further investigate the internal consistency and the
compatibility with other data sets of the ATLAS 13~TeV $Z$-boson
transverse-momentum
measurement~\cite{ATLAS:2019zci}. To this purpose, we perform an
additional fit, called 13RESwg, in which we try to improve the agreement with
this data set as much as possible, and which is otherwise equivalent to the
13RES fit. We do this by assigning the data set a weight large enough for it
to carry approximately the same weight as the remainder of the data set.
Specifically, we follow the procedure described in Sect.~4.2.3
of~\cite{NNPDF:2021njg}, which consists in assigning the data set a weight
$w=N_{\rm dat}/N_{\rm dat}^{\rm data set}$, where $N_{\rm dat}$ is the total number
of data points in the fit, and $N_{\rm dat}^{\rm data set}$ is the number of data
points in the specific data set. Such a weight multiplies the contribution of
the data set to the optimised figure of merit.

\begin{figure}[!t]
  \centering
  \includegraphics[width=0.48\textwidth]{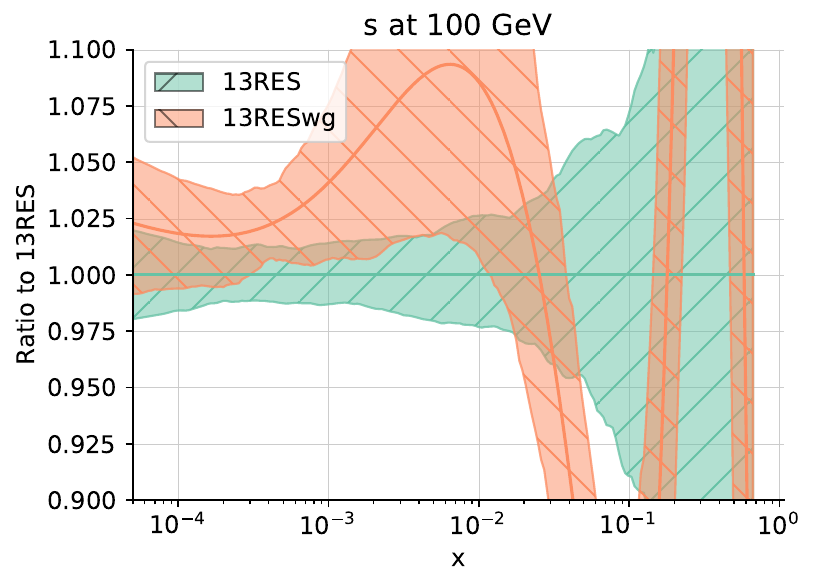}
  \includegraphics[width=0.48\textwidth]{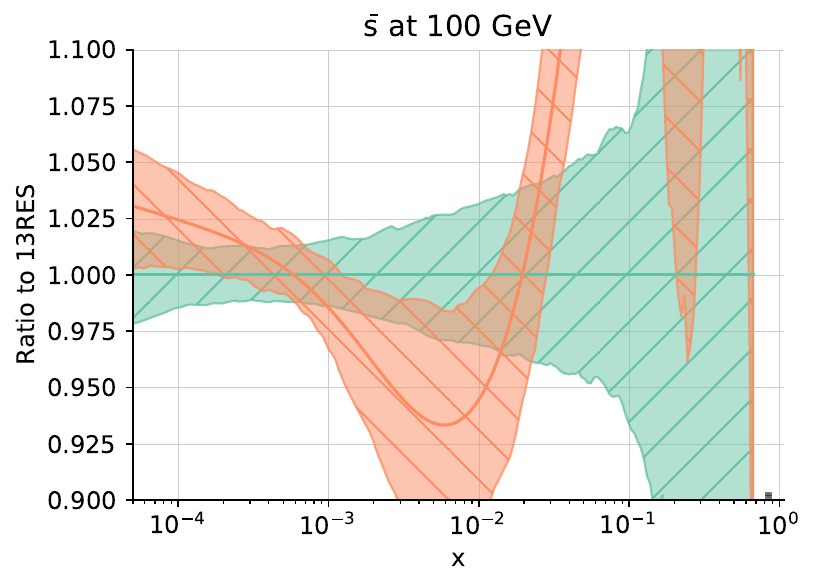}\\
  \includegraphics[width=0.48\textwidth]{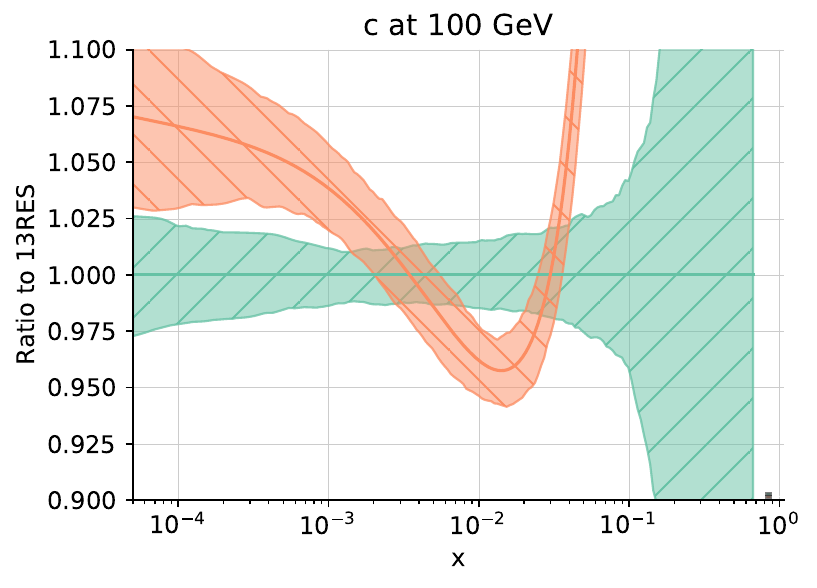}
  \includegraphics[width=0.48\textwidth]{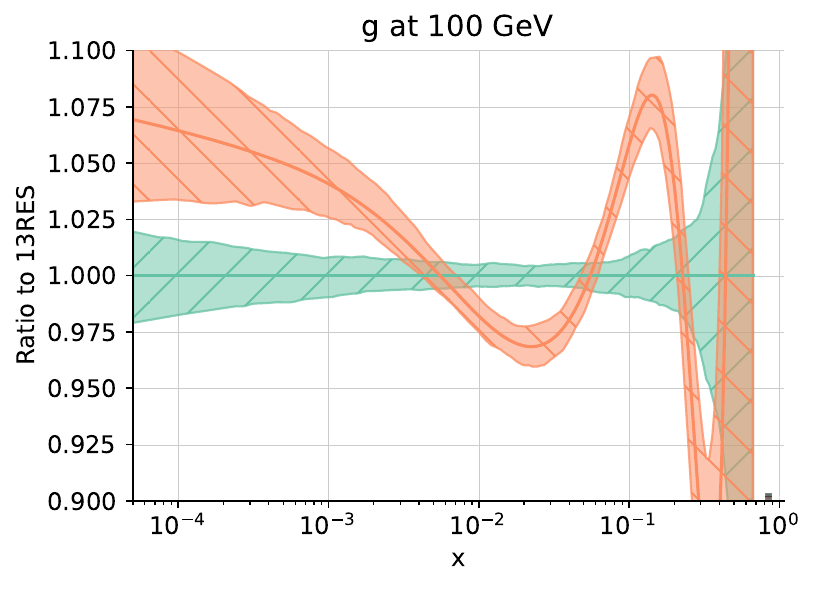}\\
  \caption{Comparison of the strange quark and anti-quark, charm quark, and
    gluon PDFs obtained from the 13RES and 13RESwg. The format is as in
    Fig.~\ref{fig:PDFs_8TeV_baseline}.}
  \label{fig:weighted_fit}
\end{figure}

We then inspect the fit quality and the resulting PDFs. Several outcomes are
possible, depending on how the $\csq$ of the weighted data set and the global
$\csq$ of the fit change, and on whether the PDFs are modified relative to
the fit in which no data set is weighted.
In Table~\ref{tab:weighted_fit}, we report
the $\csq$ values of the 13RESwg fit in the same format as
Table~\ref{tab:chi2_8TeV_baseline}, and compare them to those of the 13 RES
fit. In Fig.~\ref{fig:weighted_fit}, we display the strange quark and
anti-quark, charm quark, and gluon PDFs obtained from the 13RES and 13RESwg
fits in the same format as Fig.~\ref{fig:PDFs_8TeV_baseline}. 

%-------------------------------------------------------------------------------
\begin{table}[!t]
  \centering
  \footnotesize
  \renewcommand{\arraystretch}{1.2}
  \begin{tabularx}{\textwidth}{Xrcccccc}
    Data set
    & $N_{\rm dat}$
    & \rotatebox{90}{13RES}
    & \rotatebox{90}{13RESwg}
    \\
    \toprule
    DIS NC
    & 2100 & 1.20 & 1.30  \\
    DIS CC
    &  989 & 0.89 & 1.34  \\
    DY NC
    &  671 & 1.25 & 1.66  \\
    \quad ATLAS $\ptll$ 8~TeV ($\ptll,\yll$)
    &   44 & 1.21 & 1.33  \\
    \quad ATLAS $\ptll$ 8~TeV ($\ptll,\mll$)
    &   48 & 1.62 & 1.89 \\
    \quad CMS $\ptll$ 8~TeV ($\ptll,\yll$)
    &   28 & 1.69 & 1.51 \\
    \quad ATLAS $\ptll$ 13~TeV
    &   19 & 3.94 & 3.57 \\
    \quad CMS $\ptll$ 13~TeV
    &   16 & 0.74 & 0.88  \\
    DY CC
    &  157 & 1.06 & 1.62 \\
    top-quark pair
    &   64 & 1.02 & 1.08 \\
    single-inclusive jets
    &  356 & 0.89 & 1.45 \\
    dijets
    &  144 & 1.51 & 1.70 \\
    photon
    &   53 & 0.81 & 0.74\\
    single top-quark
    &   17 & 0.17 & 0.70\\
    \midrule
    Total
    & 4551 & 1.13 & 1.42\\
    \bottomrule
  \end{tabularx}
  \caption{The values of the $\csq$ per number of data points $N_{\rm dat}$
    for the 13RES, and 13RESwg fits. The format is the same as that of
    Table~\ref{tab:chi2_8TeV_baseline}.}
  \label{tab:weighted_fit}
\end{table}
%-------------------------------------------------------------------------------

%-------------------------------------------------------------------------------
%-------------------------------------------------------------------------------

From Table~\ref{tab:weighted_fit} and Fig.~\ref{fig:weighted_fit}, we observe
that, despite assigning a large weight to the ATLAS 13~TeV $Z$-boson
transverse-momentum data set, its $\csq$ improves only marginally, decreasing
from 3.94 to 3.57. With $N_{\rm dat}=19$ data points, this
improvement corresponds to slightly more than one standard deviation of the
$\csq$ distribution. A significant deterioration of the $\csq$ for the other
data sets is observed at the same time, in particular for deep-inelastic
scattering, charged-current DY and single-inclusive jet measurements. The
deterioration concerns also other $Z$-boson transverse-momentum measurements.
Overall, the global $\csq$ increases from 1.13 to 1.42, which, for
$N_{\rm dat}=4551$ data points, corresponds to more than 10 standard deviations
of the $\csq$ distribution. This deterioration is seen on PDFs, which display
completely different features in comparison to the 13RES fit. We have
explicitly checked that this behaviour is not
a consequence of an underlearnt fit. Following the discussion
in~\cite{NNPDF:2021njg}, we therefore conclude that the ATLAS 13~TeV $Z$-boson
transverse-momentum measurement may be affected by an overly aggressive estimate
of the experimental uncertainties, limiting its compatibility with both
similar measurements and the majority of the other data sets included in the
global fit.